\documentclass[article]{IEEEtran}

\IEEEoverridecommandlockouts
\usepackage[utf8]{inputenc}
\usepackage{graphicx}
\usepackage{amstext}
\usepackage{amsmath}
\usepackage{amsfonts,amssymb}
\usepackage{amsmath,tabu}
\usepackage{amssymb}
\usepackage[dvips]{epsfig}
\usepackage{epsfig}
\usepackage[dvips]{graphicx}
\usepackage{multirow}
\usepackage{multirow} 
\usepackage[utf8]{inputenc}
\usepackage[english]{babel}
\usepackage{caption}
\usepackage{subcaption}
\usepackage{float}
\usepackage{booktabs}
\usepackage[nospace,noadjust]{cite}
\usepackage{algorithm,algorithmic}
\usepackage{fancyhdr}
\usepackage{marginnote}

\def\bvarphi{{\boldsymbol{\varphi}}}

\def\b1{{\boldsymbol{1}}}
\def\c1{{\textcircled{a}}}

\def\bb{{\boldsymbol{b}}}

\def\bn{{\boldsymbol{n}}}

\def\bp{{\mathbf{p}}}
\def\bq{{\boldsymbol{q}}}

\def\bs{{\boldsymbol{s}}}

\def\bx{{\boldsymbol{x}}}

\def\bz{{\boldsymbol{z}}}
\def\bA{{\mathbf{A}}}

\def\bC{{\boldsymbol{C}}}

\def\bF{{\mathbf{F}}}

\def\bH{{\boldsymbol{H}}}
\def\bI{{\mathbf{I}}}

\def\bS{{\boldsymbol{S}}}

\newtheorem{theorem}{Theorem}[section]

\newtheorem{lemma}[theorem]{Lemma}
\newtheorem{prop}{Proposition}[section]
\newtheorem{ass}{Assumption}[section]

\newcommand{\mrpar}[2]{#2}
\usepackage{xcolor}
\def\BibTeX{{\rm B\kern-.05em{\sc i\kern-.025em b}\kern-.08em
    T\kern-.1667em\lower.7ex\hbox{E}\kern-.125emX}}

\begin{document}
\bstctlcite{IEEEexample:BSTcontrol}
\title{Multi-Scale Energy (MuSE) framework for inverse problems in imaging %with application in Magnetic Resonance Imaging
}
\author{Jyothi Rikhab Chand, Mathews Jacob
\thanks{The authors are with the Department of Electrical and Computer
Engineering, University of Iowa, Iowa City, IA 52242 USA (e-mail:
jyothi-rikhabchand@uiowa.edu; mathews-jacob@uiowa.edu).
This work is supported by NIH R01AG067078, R01 EB019961, and R01 EB031169.}}

\maketitle

\begin{abstract}
We introduce multi-scale energy models to learn the prior distribution of images, which can be used in inverse problems to derive the Maximum A Posteriori (MAP) estimate and to sample from the posterior distribution. Compared to the traditional single-scale energy models, the multi-scale strategy improves the estimation accuracy and convergence of the MAP algorithm, even when it is initialized far away from the solution. We propose two kinds of multi-scale strategies: a) the explicit (e-MuSE) framework, where we use a sequence of explicit energies, each corresponding to a smooth approximation of the original negative log-prior, and b) the implicit (i-MuSE), where we rely on a single energy function whose gradients at different scales closely match the corresponding e-MuSE gradients. Although both schemes improve convergence and accuracy, the e-MuSE MAP solution depends on the scheduling strategy, including the choice of intermediate scales and exit conditions. In contrast, the \mrpar{R2.C1}{i-MuSE} formulation is significantly simpler, resulting in faster convergence and improved performance. We compare the performance of the proposed MuSE models in the context of Magnetic Resonance (MR) image recovery. 
The results demonstrate that the multi-scale framework yields a MAP reconstruction comparable in quality to the End-to-End (E2E) trained models, while being \mrpar{R4.C6}{relatively unaffected by the changes in the forward model}. In addition, the i-MuSE scheme also allows the generation of samples from the posterior distribution, enabling us to estimate the uncertainty maps. 
\end{abstract}


\section{Introduction}

Maximum A Posteriori (MAP) algorithms, which formulate image recovery as the maximization of the log-posterior, are widely used in imaging. %A well-studied problem is the recovery of the Magnetic Resonance (MR) image from its noisy and undersampled Fourier measurements. 
Several efficient optimization algorithms are available to maximize a cost function involving the sum of the log-likelihood, which is a measure of Data Consistency (DC) based on a forward model of the imaging device, and the log-prior term which serves as a regularizer. Classical priors include the Tikhonov regularizer \cite{tikhonov}, total variation,  and $\ell_{1}$-penalty \cite{donoho,candes}. %In the latter case, where the regularizer is non-smooth, the proximal algorithm is commonly used, which typically alternates between the Data-Consistency (DC) update and the proximal mapping step \cite{fista}.
Several model-based deep learning algorithms, which use neural network priors, were recently introduced. For example, the Plug-and-Play (PnP) methods \cite{pnpbouman,consensus_equilibrium,rizwan_review}
replace the proximal map in the above algorithms with a pre-learned Convolutional Neural Network (CNN)-based denoiser. These algorithms offer good performance and have guaranteed convergence, stability, and robustness \cite{mukherjee2023learned,bekreview}, when the CNN is constrained as a contraction \cite{pnp_ista,mol}. % using spectral normalization of the individual layers, which  . 
The use of the above contraction constraint can restrict the achievable performance \cite{mol}. Some studies report good empirical performance even when the above constraint is not satisfied, while the algorithm may diverge in challenging cases, as seen in our experiments and in \cite{potential}. End-to-End (E2E) training strategies that rely on loop unrolling \cite{lista,admmnet,variationalnet,modl,cinenet} offer improved empirical performance than PnP methods. %However, the performance of the unrolled algorithms degrades when the forward model differs from the specific training setting. In addition, the memory demand of unrolled algorithms during training also restricts their usage in large-scale (e.g., 3D/4D) problems. 
Most of the above deep learning-based MAP approaches are not associated with an explicit cost function. \mrpar{R1.C3}{The Regularization by Denoising (RED) algorithm is an exception, which uses an explicit model for the log-prior \cite{red,red_interp}}. 

\mrpar{R1.C2}{Several deep generative frameworks were introduced in computer vision to sample from the prior distribution \cite{vae,flow,autoregressive}. Many of these schemes (e.g., variational autoencoders, flow models, autoregressive models) rely on constrained architectures to obtain an explicit and tractable likelihood. For example, flow models are based on a specialized invertible neural network, while auto-regressive models assume that the distribution is factorizable as the product of conditional distributions. Recent diffusion models \cite{yansong} rely on an implicit approach, where the derivative of the negative log-probability distribution (a.k.a score) at different noise scales is learned by a CNN. These methods rely on a multi-scale strategy to improve convergence and image quality. }However, the multi-scale strategy makes the result dependent on the specific sequence schedule, including the time-step schedule and the architecture of the deep learning model in each scale \cite{liu2023oms}.  \mrpar{R3.C4}{Non-stochastic variants \cite{ddim,consistency} and distillation strategies \cite{distillation} were recently introduced to reduce the computational complexity of diffusion algorithms}. 

The above generative models have been adapted for inverse problems, where they offer Bayesian sampling from the posterior. For example, diffusion models and \mrpar{R3.C4}{PnP methods} combine the gradient of the log-likelihood term with the score to guide image generation \cite{dps,dong_diffusion,tamir_diffusion,PnP_sampling,manifold_constraint, song_mri}. A difficulty in using these multi-scale models as principled priors in MAP estimation is that they do not provide a direct measure of the log-prior distribution;
\mrpar{R3.C4}{the evaluation involves the solution of an Ordinary Differential Equation (ODE) involving all the scales \cite{mlscoretraining, diffusion_prior}, which is computationally demanding}. Recent approaches that approximate the posterior (using a point model, Gaussian function \cite{graikos2022diffusion}, or a separate RealNVP \cite{realnvp,diffusion_prior}) utilize the ODE estimate of the prior. A challenge with the above Bayesian approaches in multidimensional applications (e.g., medical imaging) is the high computational complexity in evaluating the gradient of the log-likelihood term (e.g., multichannel nonuniform Fourier transform in MR imaging). Bayesian schemes often require the generation of a large batch of samples, followed by the derivation of the Minimum Mean Square Error (MMSE) and variance estimates. The repeated use of the log-likelihood term for the iterative computation of the samples will translate to high computational complexity. A simpler MAP estimate may often be sufficient for inference in many time-critical applications (e.g., MRI), while Bayesian estimation may be used during experiment design (e.g., choice of acceleration rate, sampling pattern).

The main focus of this work is to introduce a multi-scale CNN-based energy model for MAP estimation. {Energy-Based Models (EBMs) explicitly represent the unnormalized probability distribution using an arbitrary CNN \cite{ebm,ebm_score_energy}, which is far less restrictive than the other explicit prior models described above \cite{vae,flow,autoregressive}. Since the gradient of the negative log-probability is the score, this approach is closely related to diffusion models. However, unlike diffusion models, EBMs offer a direct measure of the log-prior (upto a constant) and therefore are ideally suited for MAP estimation.} In particular, it allows the use of conventional optimization algorithms with guaranteed convergence \cite{potential,gradientstep} for inference without requiring the CNN to be a contraction, as required by PnP methods. The main novelties of the proposed scheme are described below:   
\begin{itemize}
    \item We introduce an implicit Multi-Scale Energy model termed i-MuSE for MAP estimation. While multi-scale strategies are available for Bayesian sampling \cite{yangsongmultiscale, msdm, yansong, dps,dong_diffusion,tamir_diffusion,PnP_sampling,manifold_constraint, song_mri}, such energy-based models have not been explored in the MAP setting. In addition, using a single implicit objective function makes the framework more intuitive. 
    \item 
   Single-scale PnP score models, which have theoretical convergence guarantees, are used extensively
for MAP estimation. While multi-scale PnP score models \cite{zhang2021plug} offer better results, their
convergence properties have not been studied. In addition, current convergence guarantees for
single-scale PnP methods require restrictive contraction constraints on the CNN, which often
translate to poor performance. The proposed i-MuSE formulation enables the development of
fast optimization algorithms which, as shown in Lemma \ref{l2}, are guaranteed to converge to a stationary point of the
log-posterior without restrictive constraints on the CNN.
    \item We offer an intuitive explanation of the i-MuSE energy in Proposition \ref{lemma2}, which shows that the energy of an image is the distance of the image from the data manifold.
\end{itemize}

We learn the EBM parameters using Denoising Score Matching (DSM) \cite{vincent2010}. 
We note that the learned energy function is a smoothed version of the true prior distribution, where the variance of the Gaussian smoothing kernel is the noise variance used in DSM. This approach is associated with a trade-off between good accuracy (small variance) and good convergence of the resulting MAP algorithm (high variance). %Specifically, the learned energy is more accurate when the noise variance is small, but often has flatter regions away from the minima. This translates to poor convergence when the MAP/sampling algorithm is not initialized close to the minimum. We now describe the main contributions of the paper in the following paragraphs.
%\begin{enumerate}
%    \item 
Motivated by diffusion models \cite{yansong}, we introduce a multi-scale energy model for MAP estimation. In particular, we consider a sequence of smooth approximations of the prior, where the solution at a specific scale is used to initialize the algorithm at the next finer scale. We term this approach as an explicit Multi-Scale Energy (e-MuSE) formulation. The sequential e-MuSE optimization scheme makes the analysis of this framework challenging. Similar to diffusion models \cite{liu2023oms,wang2023learning}, the performance of this approach may depend on the
number of iterations and model choices on each scale. We hence propose an implicit Multi-Scale Energy (i-MuSE) scheme, where we learn a single energy function using DSM. The energy is learned such that its gradient evaluated at a noise-perturbed sample with a specific noise variance matches the gradient of the e-MuSE at that scale; each level set of the i-MuSE corresponds to a different scale. More importantly, we show that the i-MuSE energy of an image is the square of the distance of the image from the data manifold, which is also verified empirically. We introduce an Majorization Minimization (MM)-based algorithm, which is guaranteed to converge to a stationary point of the MAP cost function, and does not have additional hyper parameters associated with different scales. 
The smooth nature of \mrpar{R2.C1}{i-MuSE }energy offers convergence even when initialized far from the manifold. At the same time, i-MuSE closely approximates the negative log-prior near the manifold, ensuring convergence to the MAP estimate. Our experiments show that the i-MuSE scheme offers improved convergence and, in some cases, better performance than the e-MuSE scheme. 
Motivated by \cite{dps}, we introduced a modification of the DC term to further improve the convergence of i-MuSE, when the image is initialized far from the manifold. \mrpar{R1.C1, R3.C1}{We note that the i-MuSE framework has similarities to \cite{msdm}, where the prior distribution is sampled using the annealed Langevin sampling algorithm. By contrast, the main focus of our work is on using the multi-scale model in the MAP setting using faster optimization algorithms. }
%\end{enumerate}

While our focus in our work is on MAP estimation, the proposed i-MuSE model can also be used to sample from the prior as well as posterior. The samples can be obtained from the posterior distribution using the annealed Langevin Markov Chain Monte Carlo (MCMC) sampling algorithm. Using a single energy function eliminates the need to optimize the scheduling steps, thus offering rapid convergence. The presence of an explicit energy function may also enable the usage of a faster Metropolis-Hastings MCMC sampling approach.
\section{Single-scale energy model}
In this section, we first model the negative logarithmic prior distribution using a CNN denoted by $\mathcal E_{\theta}(\boldsymbol{x}):\mathbb{C}^{m} \rightarrow \mathbb{R}^+$ \cite{ebm_score_energy,potential,gradientstep}. We explain this approach in this section, before extending it to the multi-scale setting in the following sections. 
\subsection{Problem Formulation}
Let us consider the problem of recovering an image $\bx \in \mathbb{C}^{m}$ from its corrupted linear measurements $\bb \in \mathbb{C}^{n}$:
\begin{equation}\label{eq:0}
   \bA\bx+\bn = \bb
\end{equation}
where $\bA \in \mathbb{C}^{n \times m} (n \leq m) $ is a known linear transformation and $\bn \mrpar{R4.M1}{\sim} \mathcal{N}(0,\eta^{2}\bI)$ is the additive complex white Gaussian noise. Then the MAP estimate of $\bx$ from the measurements $\bb$ is given by : 
\begin{equation}
	\bx^* = \arg \max_\bx {\log p(\bx|\bb)}
\end{equation}
where $p(\bx|\bb)$ denotes the posterior distribution. Using the Bayes rule, the above cost function can be simplified as follows:
\begin{equation}\label{eq:1}
	\bx^* = \arg \min_\bx \underbrace{\dfrac{1}{2\eta^{2}} \|\bA\bx-\bb\|_{2}^2}_{-\log p(\bb|\bx)} -\log p(\bx)
\end{equation}
where $p(\bb|\bx)$ and $p(\bx)$ denotes the likelihood and the prior distribution of the unknown image, respectively. We model the prior distribution $p(\bx)$ as follows:
\begin{equation}\label{prior_new}
  p_{\theta}(\bx) = \dfrac{1}{Z_\theta}\exp\left(\dfrac{-\mathcal E_{\theta}(\bx)}{\sigma^{2}}\right),
\end{equation}
where $\sigma^2 >0$ is a parameter that is used in DSM which is described in Sec. \ref{DSM}, \mrpar{R4.C1}{$\theta$ denotes the parameters of the energy function $\mathcal E_\theta(\bx)$ that is modeled by a CNN, and $Z_{\theta}$ is a normalization constant to ensure that: 
\begin{equation}
    \int_{\bx \in \mathcal M} p_{\theta}(\bx) d\bx =1,
\end{equation}}
where $\mathcal M$ represents the manifold. The energy gradient or score $\bH_\theta (\bx)$ can be calculated using the chain rule or using PyTorch's built-in \emph{autograd} function.  
%where $\bH(\bx)=\nabla_{\bx}\left({-\log p_{\theta}(\bx)}\right)$. 
%Similar to [], we realize the above gradient using the chain rule. We provide an example in Fig. \ref{model}. A two-layered convolutional and transposed convolutional network is used to realize $\psi_{\theta}(\bx)$ and its corresponding gradient $\nabla_\bx(\psi_{\theta}(\bx))$. Note that both the networks share the weights. The output of $\psi_{\theta}(\bx)$ is used to realize the scalar energy function $-\log p(\bx)$, and the output of the decoder is used to compute the score. The resulting score will be constrained to be a conservative vector field, which ensures that it can be interpreted as the derivative of a well-defined energy function.

\subsection{Learning single-scale energy using DSM}\label{DSM}
We determine the optimal parameters of the energy function by minimizing the Fisher divergence between the true distribution $p(\bx)$ and the model distribution $p_\theta(\bx)$. This involves minimizing the objective $\mathbb E_{p (\bx)} \| \nabla_{\bx} \log p_\theta(\bx) -\nabla_{\bx} \log p(\bx) \|_{2}^2.$ Because the true distribution is unknown, we adopt the DSM technique that estimates the score of the smoothed distribution $p_{\sigma}(\tilde{\bx})$ defined as \cite{vincent2010} : 
\begin{equation}
\label{smoothed}
p_{\sigma}(\tilde{\bx}) = p(\bx) \ast\mathcal{N}(0,\sigma^{2}\bI)
\end{equation}
or equivalently $\tilde{\bx} = \bx + \sigma \bz$ where $\bz \sim \mathcal{N}(0,\bI)$. Note that $p_{\sigma}(\tilde{\bx})\rightarrow p(\bx)$ as $\sigma\rightarrow 0$. Therefore, the Fisher divergence or score-matching objective is specified by \cite{vincent2010}:
%\begin{equation}
 %L_{\sigma}(\theta)=   \mathbb E_{p_{\sigma}(\tilde{\bx})} \| \nabla_{\tilde{\bx}} \log p_\theta(\tilde{\bx}) - \nabla_{ \tilde{\bx}} \log p_\sigma(\tilde{\bx})\|_{2}^2.
%\end{equation}
%The optimization of $\theta$ using this metric is shown to be equivalent to  
\begin{eqnarray}\nonumber
 \theta^*= \arg\min_{\theta}  \mathbb E_{p(\bx)}\mathbb E_{p_\sigma(\tilde \bx|\bx)}    \| \nabla_{\tilde{\bx}} \log p_\theta(\tilde{\bx}) - \nabla_{ \tilde{\bx}} \log p_\sigma(\tilde{\bx}|\bx)\|_{2}^2\label{dsm}
\end{eqnarray}
From the definition of $p_\theta(\bx)$ in \eqref{prior_new} and using the property $\nabla_{\tilde \bx} \log p_{\sigma}(\tilde \bx|\bx) = \frac{1}{\sigma^2} (\bx-\tilde \bx)=  -\frac{\bz}{\sigma}$, the above cost function is simplified as: \begin{eqnarray}\nonumber
 \theta^{*} 
&=& \arg \min_{\theta} \mathbb E_{\bx}  \mathbb E_{\tilde {\bx}|\bx}  \left\|-\frac{\bH_{\theta} (\tilde{\bx})}{\sigma^2} + \frac{\bz}{\sigma}  \right\|_{2}^2\\\label{dsm}
&=& \arg \min_{\theta} \mathbb E_{\bx} ~\mathbb E_{\bz}  \left\|\bH_{\theta} ({\bx}+\sigma \bz) - \sigma \bz  \right\|_{2}^2\label{dsm}
\end{eqnarray}
\mrpar{R4.C5}{We note that the computation of $\bH_\theta$ does not involve the computation of the normalization constant $Z_\theta$, because it is independent of $\bx$. This makes the approach insensitive to the exact constant, which is difficult to evaluate.}
We note that DSM training is used in PnP methods and diffusion models. The main distinction from these settings is that $\bH_{\theta}$ is constrained as the gradient vector field $\bH_{\theta}=\nabla_\bx \mathcal{E}_{\theta}$. A major benefit of this approach is that $\mathcal{E}_{\theta}$ is an explicit model for the \mrpar{R4.M2}{negative} log-prior, which enables the direct evaluation of the log-prior for an arbitrary image. When only the score function is available, diffusion models must evaluate the log-prior using numerical integration \cite{yansong,diffusion_prior}:
\begin{equation}\label{LineInt_diffusion}
    \log p_\theta(\bx) = \int_{0}^{T}\nabla .{s}_\theta (\bx(t),t) dt+ \log p_\pi (\bx_T) 
\end{equation}
involving all the scales, where \mrpar{R4.M3}{$\log p_\pi (\bx_T) $} denotes the final Gaussian noise distribution in diffusion models, $T$ is a constant, ${s}_\theta (\bx(t),t)$ represents the score function of the diffusion model at scale $t$, and \mrpar{R2.C3}{$\nabla .$ represents the divergence operator}. Since this integral computation is expensive, it is impractical to evaluate it during each iteration of the inference.

%\textcolor{red}{For simplicity, we omit the dependence of a network $E_{\theta}$ on its parameters $\theta$, unless necessary (in the training steps). In the DSM setting with $\sigma\neq 0$, we consider a scaled version of the energy $Q_{\sigma}(\bx) = \frac{1}{\sigma^2} E(\bx) $ and $F_{\sigma} = \nabla Q_{\sigma}=\frac{1}{\sigma^2}H(\bx)$, which will simplify the downstream analysis.}

\subsection{MAP image recovery using single-scale energy}\label{mm}
Once the optimal parameters of the prior distribution $p_\theta(\bx)$ are estimated, the MAP estimate of the unknown image $\bx$ is obtained by solving:
\mrpar{R1.C5}{\begin{eqnarray}
\label{eq:P1}
	\bx^* &=& \arg \min_\bx f_{\rm MAP}(\bx)\\\nonumber
 &=& \arg \min_\bx\dfrac{1}{2\eta^{2}} \|\bA\bx-\bb\|_{2}^2 + \dfrac{\mathcal{E}_\theta(\bx)}{\sigma^{2}}
\end{eqnarray}}
We optimize the above problem using the MM framework which minimizes the problem (\ref{eq:P1}) iteratively. At each iteration, the MM scheme derives a surrogate function $g(\bx|\bx_{n})$ such that $f_{\rm MAP}(\bx)\leq g(\bx|\bx_{n})$, which is minimized to obtain the next iterate. To construct a surrogate function for the cost function in (\ref{eq:P1}), we use the following lemma.
\begin{lemma}
\label{lemma0}
Let $\mathcal E_\theta(\boldsymbol x)$  be a continuously differentiable function with a Lipschitz constant $L$ (we discuss in Sec. \ref{map_exps}.B regarding the Lipschitz value estimation). Then, $f_{\rm MAP}(\bx)\leq g(\bx|\bx_{n})$, where $g(\bx|\bx_{n})$ is the quadratic function:
\begin{eqnarray}\nonumber
    g(\bx|\bx_{n}) &=& \dfrac{\|\bA\bx - \bb\|^{2}}{2\eta^{2}}+ \dfrac{\mathcal E_{\theta}(\bx_{n})}{\sigma^{2}}+\dfrac{L }{2\sigma^2}\|\bx-\bx_{n}\|^{2}+\\\label{surrogate}
    &&\qquad\qquad\dfrac{\textrm{Re}\left(\boldsymbol H_{\theta}(\bx_{n})^{H}(\bx - \bx_{n})\right)}{\sigma^{2}} 
    \vspace{-2mm}
\end{eqnarray}
\end{lemma}
\begin{IEEEproof}
Please refer Lemma 12 in \cite{mm_tutorial}. \mrpar{R1.C7}{While the above theorem assumes that the energy function is continuously differentiable, the network that we implemented uses ReLU activation function that is not continuously differentiable. We note that MM algorithm offers convergence in our context, even though the assumption is violated in the strict sense. }

\end{IEEEproof}
The minimization of the above surrogate function offers a closed-form solution given as \footnote{\mrpar{R2.C2}{Wirtinger calculus is applied for the DC term when the image is complex to compute the gradient.}}:
\begin{equation}\label{mm_up}
        \bx_{n+1} = \left(\dfrac{\bA^{H}\bA}{\eta^{2}}+\dfrac{L}{\sigma^{2}}\bI\right)^{-1}\left(\dfrac{\bA^{H}\bb}{\eta^{2}} + \dfrac{L \bx_{n} - \boldsymbol H_{\theta}(\bx_{n})}{\sigma^{2}}\right)
\end{equation}   
For a large-dimensional $\bx$, inverting the $\bA$ operator is computationally expensive. Therefore, we adopt the conjugate gradient algorithm to obtain the next iterate $\bx_{n+1}$. 
\section{Multi-scale energy (MuSE) models}
As discussed in Sec. \ref{DSM}, the energy function estimated using DSM corresponds to a smooth approximation of the log prior (see \eqref{smoothed}), which depends on the noise variance $\sigma$ used in DSM. Small values of $\sigma$ translate to a more accurate estimate of the distribution. However,  when the data is localized to a low-dimensional manifold, the  learned energy may have flatter regions with low gradients away from the manifold. This affects the convergence of the algorithm, when it is initialized far from the data manifold. Inspired by continuation strategies used in non-convex optimization algorithms, we introduce a similar approach to improve the convergence. In the next section, we propose an explicit multi-scale strategy to improve the trade-off between accuracy and convergence. Subsequently, we introduce a more efficient implicit scheme to further improve the performance and convergence rate.

\subsection{Explicit Multi-Scale Energy  (e-MuSE) model}\label{sec_emuse}
%\subsection{Learning multi-scale (explicit) energy using DSM}
We consider a sequence of energies $\mathcal{E}_{\theta(\sigma)}$, each of which are learned at a different scale $\sigma$:
\begin{eqnarray}\nonumber\label{explicit}
 \theta^*(\sigma) 
&=& \arg \min_{\theta}  \mathbb E_{\bx} ~\mathbb E_{\bz}  \left\|\nabla_\bx \mathcal{E}_{\theta(\sigma)} (\bx+\sigma \bz) - \sigma \bz  \right\|_{2}^2\\
 \end{eqnarray}
The energy $\mathcal{E}_{\theta(\sigma)}$ correspond to smooth approximations of the negative log-prior at different scales, parameterized by the scale parameter $\sigma$. The MAP objective at the scale $\sigma$ is thus given by:
\begin{equation}\label{eq:P2}
f_{_{\sigma}}(\bx)= \dfrac{\|\bA\bx-\bb\|_{2}^2}{2\eta^{2}}  + \dfrac{\mathcal{E}_{\theta(\sigma)}(\bx)}{\sigma^2}
\end{equation}
%\subsection{MAP image recovery using e-MuSE}\label{emuse_algo}
For a fixed $\sigma$, we minimize \eqref{eq:P2} using the MM algorithm discussed in Sec. \ref{mm}. We start with a coarse scale (large $\sigma$), where the energy function is smooth and has well-defined gradients even at locations that are far from the data manifold. Once the algorithm has converged, we use the solution to initialize the problem with the next smaller $\sigma$. 
We gradually reduce the scale parameter to get a sequence of solutions. We term this multi-scale algorithm as e-MuSE. The pseudocode of e-MuSE is shown in Algorithm 1. 

We note that annealing strategies, where the networks are conditioned on the noise variance, have been introduced in PnP methods \cite{zhang2021plug}. However, these methods have not been explored in the context of explicit energy-based models, which can also be used to sample from the posterior distribution. 

%Plug-and-Play Image Restoration with Deep Denoiser Prior 
%FFDNet: Toward a fast and flexible solution for cnn-based image denoising 

 %\vspace{-1em}

%\begin{center}
%\begin{tabular}{@{}p{8.5cm}}
%\hline
%\hline
%\bf{Algorithm 1: Pseudocode of Gradient Descent algorithm to minimize MAP problem } \\
%\hline
%\hline
%{\bf{Input}}: Forward operator $\bA$, Pre-trained denoiser $H_{\theta}(\bx_{n})$ Noise variances - $\{\eta^{2}, \sigma^{2}\}$, Step size $\gamma = \dfrac{1}{\dfrac{1}{\eta^{2}}+\dfrac{L}{\sigma^2}}$ \\
%{\bf{Initialize}}: Set $n=0$.  Initialize $\bx_{0}$. \\ 
%{\bf{Repeat}}: Given ${{\bx}_{n}}$ perform the $n+1$-th step.\\
%\hspace{5mm} Compute $\bx_{n+1} =  \bx_{n}- \gamma\left(\dfrac{\bA^{T}(\bA\bx_{n}-\bb)}{\eta^{2}} +  \dfrac{H(\bx_{n})}{\sigma^{2}}\right)$\\
%\hspace{10mm}$n\leftarrow n+1$\\
%{\bf{until convergence}}\\
%{\bf{Output}}: $\bx_{\rm{GD}}^{*} = \bx_{n+1}$
%\\
%\hline
%\hline
%\end{tabular}
%\end{center} 

\subsection{Implicit Multi-Scale Energy  (i-MuSE) model}\label{sec_imuse}

 In the previous section, we relied on multiple models $\mathcal{E}_{\theta(\sigma)}$ for a preset sequence of \mrpar{R4.M4}{scale} parameters, denoted by $\Sigma=\{\sigma_1,...,\sigma_k\}$. Although this approach improves convergence, the final result depends on the specific sequence $\Sigma$. In addition, the optimization at each step to convergence may not be needed for generating high-quality reconstructions. Thus, a careful optimization of the hyper parameters (the set $\Sigma$ as well as exit conditions at each step) may be needed to realize a fast algorithm that can offer high quality reconstructions. The dependence of the final solution on the intermediate solutions along the specific path also makes it challenging to analyze the quality of the reconstructions. 

\vspace{-2mm}
\begin{center}
\begin{tabular}{@{}p{9cm}}
\hline
\hline
\bf{\mrpar{R5.C1}{Algorithm 1}: Pseudocode of e-MuSE} \\
\hline
\hline
{\bf{Inputs}}: Forward operator: $\bA$, inference noise variance: $\eta^{2}$, scale parameters: $\{\sigma_{i}^{2})\}_{i=1}^{k}$\\ Pre-trained energy models: $\{\mathcal{E}_{\theta_1(\sigma_{1})},\cdots,\mathcal{E}_{\theta_{k}(\sigma_{k})}\}$\\
{\bf{Initialize}}: Set $(n,k)=0$, Initialize $\tilde{\bx}_{0}$ and $\bx_{0}$.  \\
{\bf{Repeat}}: Given ${{\tilde{\bx}}_{k}}$ perform the $k+1$-th step. \\
\hspace{10mm}$n \leftarrow 0$\\
\hspace{10mm}{\bf{Repeat}}: Given ${{\bx}_{n}}$ perform the $n+1$-th step.\\
\hspace{13mm} Compute $\bx_{n+1}$ using (\ref{mm_up})\\ 
\hspace{13mm}$n\leftarrow n+1$\\
\hspace{10mm}{\bf{until convergence}}\\
\hspace{10mm}{$\tilde{\bx}_{k} = \bx_{n+1}$, $k\leftarrow k+1$, {\bf{until convergence}}}\\
{\bf{Output}}: $\bx_{\rm{e-MuSE}}^{*} = \tilde{\bx}_{k}$
\\
\hline
\hline
\end{tabular}
\end{center}

%\subsection{Learning multi-scale (implicit) energy using DSM}
To realize a simpler and more efficient algorithm, we propose to learn a single energy function $\mathcal{I}_\theta(\bx): \mathbb{C}^{m} \rightarrow \mathbb{R}^{+}$ by minimizing:
\begin{eqnarray}\label{implicit}
\theta^* 
= \arg \min_{\theta}  \mathbb E_{\sigma}~\left(\mathbb E_{\bx} ~\mathbb E_{\bz}  \left\|\nabla_\bx \mathcal{I}_{\theta} (\bx+\sigma \bz) - \sigma \bz  \right\|_{2}^2\right)
\end{eqnarray}
Since we replace a collection of networks by a single one, we  use a larger network $\mathcal{I}_{\theta}$ in this setting compared to $\mathcal{E}_{\theta(\sigma)}$  in e-MuSE. We note that the trained models $\mathcal{I}_\theta$ and $\mathcal{E}_{\theta(\sigma)}$ are related to each other as:
 \begin{equation}\label{equivalence}
     \nabla_\bx \mathcal{I}_\theta(\bx + \sigma \bz) = \nabla_\bx \mathcal{E}_{\theta(\sigma)}(\bx + \sigma \bz)
 \end{equation}
 While \mrpar{R4.M5}{$\mathcal{I}_\theta(\bx)$} does not correspond to the e-MuSE energy at any specific scale, the response of the i-MuSE score to a noisy input $\bx + \sigma \bz$  is exactly the same as the response of e-MuSE score function designed for that specific $\sigma$. We refer to the learned implicit multi-scale energy $\mathcal{I}_{\theta}$ as implicit MuSE (i-MuSE) to differentiate it from the collection of explicit models.

Because we use a single energy, the recovery of the image is obtained by optimizing a single cost function:
 \begin{eqnarray}\label{imuse}
 \bx^* &=& \arg \min_\bx \underbrace{\dfrac{1}{2\eta^{2}} \|\bA\bx-\bb\|_{2}^2 + \dfrac{\mathcal{I}_\theta(\bx)}{\sigma_f^{2}}}_{\mrpar{R4.M6}{-\log p_\theta(\bx|\bb)}}\nonumber\\
   &=& \arg \min_\bx  \dfrac{1}{2\zeta^2}\|\bA\bx-\bb\|_{2}^2 + \mathcal{I}_\theta(\bx),
\end{eqnarray}
where $\zeta=\eta/\sigma_f$ and $\sigma_f$ is the standard deviation of the finest level used to train i-MuSE. We solve the problem \eqref{imuse} using the MM algorithm discussed in Sec. \ref{mm}. In particular, we majorize \eqref{imuse} using Lemma \ref{lemma0}, and arrive at the following update equation: 
\begin{equation}\label{mm_imuse}
        \bx_{n+1} = \left(\dfrac{\bA^{H}\bA}{\zeta^{2}}+L~\bI\right)^{-1}\left(\dfrac{\bA^{H}\bb}{\zeta^{2}} + L \bx_{n} - \nabla_{\bx_{n} }\mathcal{I}_{\theta}(\bx_{n})\right).
\end{equation}   
The pseudocode of i-MuSE algorithm is shown in Algorithm 2.
 \vspace{-2mm}
\begin{center}
\begin{tabular}{@{}p{9cm}}
\hline
\hline
\bf{\mrpar{R5.C1}{Algorithm 2}: Pseudocode of i-MuSE} \\
\hline
\hline
{\bf{Inputs}}: Forward operator: $\bA$\\ Pre-trained energy model: \mrpar{R4.M5}{$\mathcal{I}_\theta(\bx)$}, $\zeta^{'}=\dfrac{\eta}{\sigma_f}$\\
{\bf{Initialize}}: Set $n=0$. Initialize $\bx_{0}$.  \\
{\bf{Repeat}}:\\
\hspace{10mm}Given ${{\bx}_{n}}$ perform the $n+1$-th step.\\
\hspace{10mm} Update $\bx_{n+1}$ using (\ref{mm_imuse}), $n\leftarrow n+1$\\
{\bf{until convergence}}\\
{\bf{Output}}: $\bx_{\rm{i-MuSE}}^{*} = {\bx}_{n+1}$
\\
\hline
\hline
\end{tabular}
\end{center}
\subsection{Proof of convergence}
We now show that the MM-based algorithm \eqref{mm_imuse}  will converge to a stationary point of the cost function $-\log p_\theta (\bx|\bb)$ defined in  \eqref{imuse}. Note that since we explicitly model the negative log-prior as a multi-scale energy model, the MM-based algorithm is guaranteed to converge, as shown below, even when the score is not constrained to be a contraction.  We would like to point that, unlike e-MuSE, which employs multiple models $\mathcal{E}_{\theta(\sigma)}$ for a preset sequence of scale parameters, i-MuSE  uses a single implicit multi-scale energy function. Consequently, we can only establish the convergence of e-MuSE to a stationary point of the optimization problem corresponding to each scale $\sigma$. Using a single energy function, as in i-MuSE, enables us to establish the convergence guarantees in a straightforward manner, as shown below:
\begin{lemma}\label{l2}
The sequence $\{\bx_{n}\}$ generated by the MM algorithm will monotonically decrease the cost function $-\log p_\theta (\bx|\bb)$ in \eqref{imuse} and will converge to a stationary point of \eqref{imuse}.
\end{lemma}
\begin{IEEEproof}
Refer \cite{mm_conv} for the proof. Note that the proof in \cite{mm_conv} is based on the following assumptions:
\begin{enumerate}
    \item Cost function $f(\bx)$ is lower bounded by a finite value. 
    \item The surrogate function satisfies $g(\bx_n|\bx_n)=f(\bx_n)$ and $g(\bx|\bx_n)\geq f(\bx)$.
\end{enumerate}
Note that since $\mathcal{I}_\theta(\bx) \geq 0; \forall \bx \in \mathbb{C}^m$, we have $-\log p_\theta (\bx|\bb)\geq 0$. Thus, the proposed algorithm converges to a stationary point of $-\log p_\theta (\bx|\bb)$ defined in \eqref{imuse}. 
\end{IEEEproof}
\subsection{Analysis of the i-MuSE energy}
 We now analyze the implicit energy in a special case where the data samples live on a smooth and low-dimensional manifold $\mathcal M$ in high-dimensional space. 
 
\begin{ass} Assume the probability of $\bx$ denoted as $Pr(\bx)$ is supported on the weakly compact set $\mathcal M;~Pr(\mathcal M_c) = 0$, where $\mathcal M_c$ denotes the complement of the set $\mathcal M$. We assume $\mathcal M \subset \mathbb C^m$ is a smooth $d<<m$ dimensional manifold.
\end{ass}
 
 Because the gradients of i-MuSE and e-MuSE match (see \eqref{equivalence}), $\mathcal{I}_\theta$ is expected to be smooth and have well-defined gradients far away from $\mathcal M$, similar to e-MuSE at the coarsest scales. Closer to $\mathcal M$, the i-MuSE function is expected to closely match e-MuSE at the finest scale, thus offering estimates that closely match the MAP estimate. The implicit multi-scale strategy may eliminate the iterations at higher scales in e-MuSE, which may not contribute to improved quality of the final estimate. Consequently, the proposed implicit approach may translate to faster convergence, compared to the successive minimization of e-MuSE objectives. 
 
 The following result offers an alternative interpretation of the single i-MuSE model.  
\begin{prop}
\label{lemma2}
    Under the manifold assumption, the energy of an arbitrary point $\tilde{\bx} \sim p_{\sigma}(\tilde{\bx})$ for an arbitrary $\sigma$ is given by: 
    \begin{equation}
        \mathcal{I}_\theta(\tilde \bx) = \frac{1}{2}~d(\tilde{\bx},\mathcal M)^2,
    \end{equation}
    where $d(\tilde{\bx},\mathcal M)$ is the Euclidean distance of $\tilde{\bx}$ to the manifold. 
\end{prop}
The proof is shown in Appendix A.
This result shows that the energy at any point $\tilde{\bx}$ is a smooth quadratic function of the distance of $\tilde{\bx}$ from $\mathcal M$, even for high values of $\sigma$. This is also verified empirically in Fig. \ref{Quadratic_illustration}. Therefore, the use of the i-MuSE model as a prior in MAP optimization yields a smooth and well-behaved optimization landscape. 

The above property also shows that the Hessian of the energy satisfies $\nabla^2_\bx \mrpar{R4.M5}{\mathcal{I}_{\theta}(\bx)} \approx \mathbf I$ away from $\mathcal M$, enabling the use of simple optimization tools to rapidly converge to the projection of an arbitrary point $\tilde{\bx}$ to $\mathcal M$ or the intersection of $\mathcal M$ and the hyperplane $\mathbf A\bx = \bb$, irrespective of initialization.

\begin{figure}
\centering
    \includegraphics[width=0.4\textwidth]{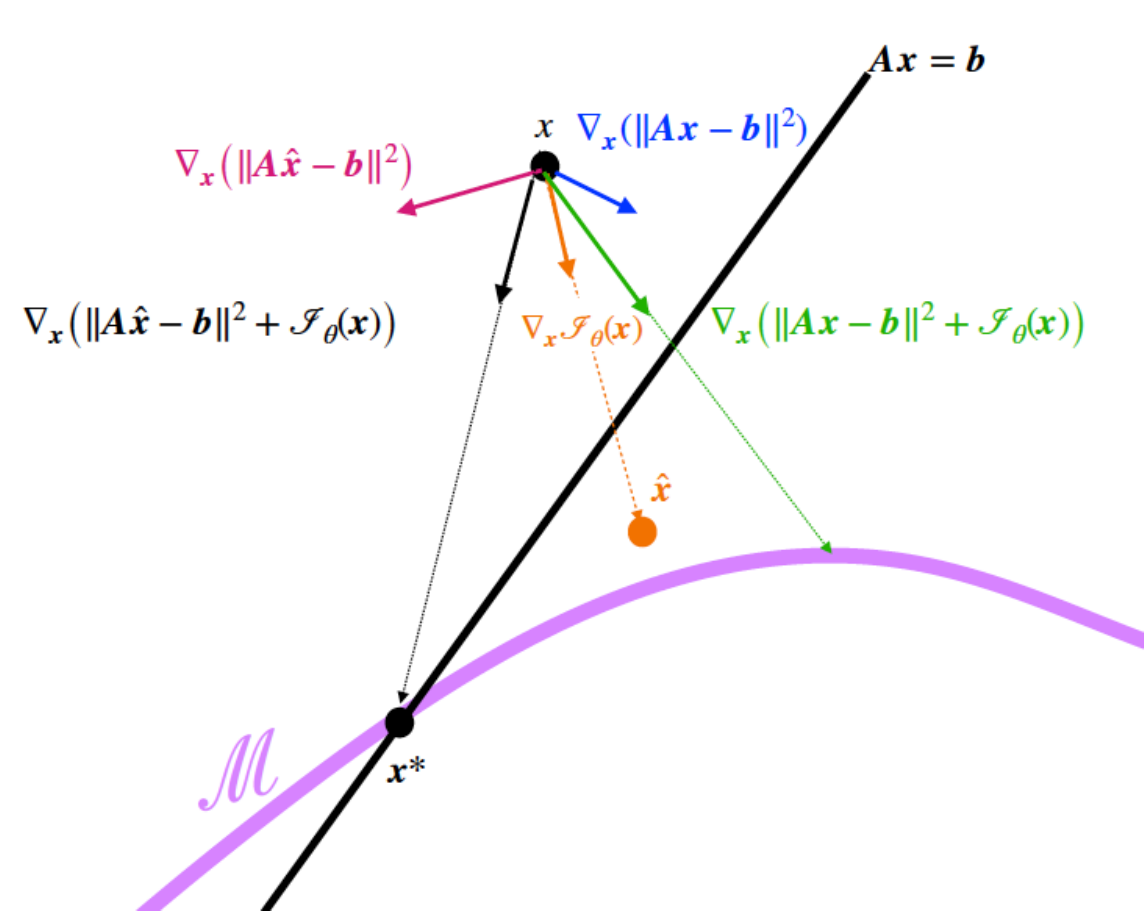}
    \caption{Illustration of improved gradient direction described in Section \ref{speedup}. The optimal solution denoted as $\bx^*$ is the intersection of $\mathcal M$ and the hyperplane $\bA\bx=\bb$. Descending along the negative gradient of the DC term alone (shown as a blue arrow) will bring it closer to the hyperplane, while descending along the negative gradient of the energy $\nabla_\bx \mathcal{I}_\theta(\bx)$ (shown as orange arrow) will bring it closer to the manifold (denoted as $\hat{\bx}$). In this example, the gradient of the combination of DC and energy terms shown in green (i.e. the gradient of \eqref{imuse}) may not point towards the optimal solution. On the contrary, the gradient of the modified cost function $f_{\rm m-MAP}(\bx)$ shown in black (which is the direction along which $\bx$ should move such that $\hat{\bx}$  satisfies the DC term) will point towards the optimal solution, thus offering a faster convergence. }
    \label{illusgrad}
\end{figure}
\subsection{Accelerating the convergence of i-MuSE}
\label{speedup}
Motivated by \cite{dps}, we introduce an improved guidance strategy to further accelerate the convergence rate of i-MuSE. 
Diffusion posterior sampling \cite{dps} uses the CNN-based score function to remove the noise in the current iterate as: 
\begin{equation}
\label{proj}
\hat{\bx} = \bx-\nabla_\bx \mathcal{I}_{\theta}(\bx),
\end{equation}
which is then used to derive the gradient of the DC term (note that \cite{dps} does not use a conservative score function). We thus use $\nabla_{\bx} \|\mathbf A\hat{\bx}-\bb\|_{2}^2$ to guide diffusion rather than $\nabla_{\bx} \|\mathbf A{\bx}-\bb\|_{2}^2$. Modifying the DC gradient as discussed above is equivalent to the following modified loss function:
 \begin{equation}\label{imusemod}
f_{\rm m-MAP}(\bx)= \dfrac{\|\bA \hat{\bx}-\bb\|_{2}^2}{2\zeta^{2}}  + \mathcal{I}_\theta(\bx)
\end{equation}
We provide an illustration in Fig. \ref{illusgrad}, which shows that the modified loss function may offer improved convergence. Furthermore, note that as $\bx\rightarrow \bx^*$, where $\bx^* \in \mathcal M$, we have $\hat{\bx} \rightarrow \bx$, and hence $\|\mathbf A\hat{\bx}-\bb\|_{2}^2 \approx \|\mathbf A\bx-\bb\|_{2}^2$ in this case. Thus, the modification of the DC term is not expected to change the final result, and only improve the convergence rate. 

We show in Appendix B that an MM-based algorithm to minimize \eqref{imusemod} is given by:
\begin{eqnarray}\label{sol_mod_dc}
        \bx_{n+1} &=& \left(\dfrac{\beta^2 \bA^T\bA}{2\zeta^2}+ {L\bI}\right)^{-1}\bq, ~~\mbox{where}\\\nonumber
        \bq &=& \dfrac{\beta^2 \bA^T\bA \bx_n-\nabla_{\bx_n} (\|\bA \widehat{\bx}_n)-\bb\|_{2}^2)}{2\zeta^2}\\&&\qquad+ L\bx_n-\nabla_{\bx_n} \mathcal{I}_\theta(\bx_n),
\end{eqnarray}
where $\beta$ is the Lipschitz constant of the mapping \eqref{proj} and its value is chosen as discussed in Sec. \ref{map_exps}.B. 

% In most cases, evaluating the gradient of the regularizer $\nabla_\bx \mathcal{I}_\theta(\bx)$ is computationally less expensive than evaluating the gradient of the data term involving Fourier transform.
% We hence consider the projection of $\bx$ onto $\mathcal M$ denoted as $P_{\mathcal M}(\bx)$, and consider the following modified loss function for faster convergence:
 % \begin{equation}\label{imusemod}
 %f_{\rm m-MAP}(\bx)= \dfrac{\|\bA P_{\mathcal M}(\bx)-\bb\|_{2}^2}{2\zeta^{2}}  + \mathcal{I}_\theta(\bx)
 %\end{equation}
 %We implement $P_{\mathcal M}(\bx)$ using gradient descent. In our experience, a single gradient descent step:
 %\begin{equation}
 %P_{\mathcal M}(\bx) \approx \bx - \nabla_\bx \mathcal{I}_\theta(\bx)
 %\end{equation}
 %is sufficient to improve the convergence rate.
\subsection{Bayes estimation}
Unlike PnP and E2E frameworks, the i-MuSE formulation provides the posterior, which can be used to sample from the distribution. Since i-MuSE involves a single energy function, samples can be generated from the posterior distribution using the Langevin MCMC algorithm \cite{ebm}:
\begin{equation}\label{Langevin}
    \bx_{n+1} = \bx_n + \epsilon~ \nabla_\bx \log p_\theta(\bx|\bb) + \sqrt{2\epsilon t }~Z_n  
\end{equation}
where $-\log p_\theta(\bx|\bb)$ is defined in \eqref{imuse}, $\epsilon>0$ is the step-size, $t>0$ is the temperature, $Z_n$ and $\bx_0 \in \mathcal{N}(0,\bI)$.  \mrpar{R2.C4}{The true posterior in \eqref{imuse} or the approximate one specified by \eqref{imusemod} may be used in the above expression. In this work, we use \eqref{imusemod} for faster convergence, where the samples correspond to the estimate of the true posterior provided by the Tweedie's formula \cite{tweedie}. However, we note that the steps are iterated, making the samples closer to that of the true posterior.} The empirical mean and variance of the samples can then be used to compute MMSE and the uncertainty of the reconstructed sample, respectively \cite{uecker}.  Because i-MuSE offers an expression for the negative log-posterior, one may use the Metropolis-Hastings algorithm with larger step sizes leading to faster convergence. However, we do not consider them in this work.

\section{Experiments: MAP estimation}\label{map_exps}
\subsection{Data set}
We evaluate the reconstruction performance of the proposed energy models in the context of \mrpar{R4.C7}{ the reconstruction of 2D MR images}. We consider multi-channel acquisition, where the sensitivity profiles are estimated using \cite{espirit}. Under this parallel imaging setting,  the forward model is specified as  $\mathbf A=(\bS\bF\bC)\bx$, where $\bS$ is the sampling matrix, and $\bF$ is the multichannel Fourier transform, and $\bC$ corresponds to the Coil Sensitivity Map (CSM). We use the publicly available parallel \cite{knee_dataset} T2-weighted brain data set, which is a twelve-channel data set and consists of complex images of size $320 \times 320$. The data set was split into $45$ training, $5$ validation, and $50$ test subjects. We scaled the k-space data such that the magnitude of the images are less than one. We evaluated the models for different acceleration factors using non-uniform Poisson variable density masks.  
\subsection{Architecture and implementation}
We now describe the architectures and pre-training procedures of the different flavors of models used in this paper. \mrpar{R2.C2}{Note that irrespective of the architecture, when the images are complex, we represent them as a two-channel image (real and imaginary channels) for the energy and score calculation.}
\subsubsection{Single-scale Energy Model (EM) and e-MuSE} 
\begin{figure}[!ht]
\centering
    \includegraphics[trim={1cm 7cm 1cm 5cm},width=1\linewidth]{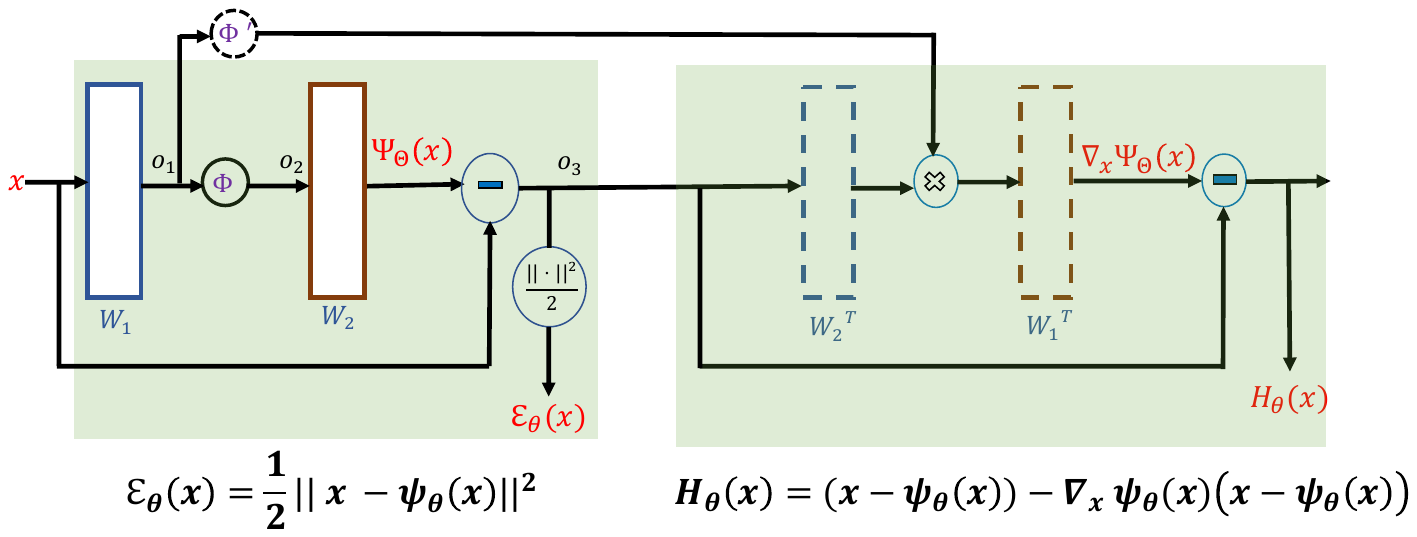}
\caption{Illustration of a two-layer network  \protect\footnotemark to realize $\mathcal{E}_{\theta}(\bx) = \dfrac{1}{2}\|\bx - \psi_{\theta}(\bx)\|^{2}$. The corresponding score $\bH_{\theta}(\bx)=\nabla_\bx \mathcal{E}_{\theta}(\bx)$ is computed using the chain rule. We note that when pooling operations are used in the energy network, the corresponding score network resembles a U-Net.}
\label{energy_models}
\end{figure}

We model the single-scale energy $\mathcal{E}_{\theta}: \mathbb C^{m} \rightarrow \mathbb R+$ as: 
\begin{eqnarray}\label{e1}
    \mathcal{E}_{\theta}(\bx) &=& \frac{1}{2}{\|\bx-{\psi}_{\theta}(\bx)\|^{2}},
\end{eqnarray}
where $\theta$ denotes the parameters of the energy function, $\psi_{\theta}(\cdot): \mathbb C^m\rightarrow \mathbb C^m$ is a CNN network. If ${\psi}_{\theta}$ can be viewed as a denoiser; therefore, the energy can be interpreted as the magnitude of noise in the image $\bx$. Fig. \ref{energy_models} illustrates a simple two-layer network to compute the energy in \eqref{e1}, where the score function is computed using a chain rule. 

We realized single-scale and e-MuSE energies for different scales using \eqref{e1}, where $\psi_{\theta}(\cdot)$ was built using a five-layer convolutional network. We used Rectified-Linear Unit (ReLU) activation for all layers, except the final one. The convolutional layer consisted of $64$ channels with a $3 \times 3$ filter size. The energy gradient or score function $\bH_\theta(\bx)$ was calculated using the built-in autograd function of PyTorch.
\footnotetext{In Fig. \ref{energy_models}, $W_1$ and $W_2$ represents convolutional layers of appropriate size; $\phi$ denotes non-linear activation function; $W_1^T$ and $W_2^T$ represents the transposed convolutional layer; and $\phi^{'}$ is the gradient of the non-linear activation function. }

We trained the single-scale EM using \eqref{dsm} to remove  zero-mean Gaussian noise with standard deviation  $\sigma= 0.01$. To implement e-MuSE five separate EM $\{\mathcal{E}_{\theta_i(\sigma_i)}(\bx)\}_{i=1}^{5}$  were trained according to \eqref{explicit} to remove five different zero-mean Gaussian noise with standard deviations $\Sigma=\{\sigma_1= 0.09, \sigma_2 = 0.07, \sigma_3 = 0.05, \sigma_4 =0.03,\sigma_5 =0.01
\}$. Each $\mathcal{E}_{\theta(\sigma)}(\bx)$ was modeled as in \eqref{e1}. \\\mrpar{R4.C3}{Note that, the single-scale EM and e-MuSE implements the update step \eqref{mm_up} for image recovery, which requires the knowledge of the Lipschitz constant $L$ of  $\mathcal{E}_{\theta(\sigma)}(\bx)$ whose value is in in-turn dependent on the CNN ${\psi}_{\theta}(\bx)$. However, determining the exact Lipschitz constant of CNN is an NP-hard problem \cite{CLIP}. In this work, we estimated the approximate Lipschitz constant using the CLIP \cite{CLIP} approach.  The approximate $L$ corresponding to each $\mathcal{E}_{\theta(\sigma)}(\bx)$  was found to be equal to $\{1.1174\: (\sigma=0.09), 1.0452 \:(\sigma=0.07), 0.9695 \:(\sigma=0.05),  0.9410\: (\sigma=0.03), 0.6903\: (\sigma=0.01)\}$. Since CLIP only gives an approximate value of the true Lipschitz, and the algorithm's convergence is guaranteed as long as the chosen $L$ value is greater than the true Lipschitz, we selected a conservative value of $L = 5$ in the experiments. }\\

\subsubsection{i-MuSE model} We used a similar formulation for i-MuSE:
\begin{equation}
    \mathcal{I}_\theta(\bx)=\dfrac{1}{2}\|\bx-\bvarphi_\theta(\bx)\|^{2}
\end{equation}
We used a deeper DRUNet \cite{zhang2021plug} to represent $\bvarphi_\theta(\bx)$ compared to $\psi_\theta(\bx)$ in \eqref{e1}. The architecture of DRUNet is similar to an U-Net with four downscaling and upscaling layers consisting of 64,
128, 256, and 512 channels. Each layer has a skip connection between $2 \times 2$ strided convolution downscaling and $2 \times 2$ transposed convolution upscaling operation. However, unlike U-Net, DRUNet has four additional residual
blocks in each downscaling and upscaling layer. The parameters of $\mathcal{I}_\theta(\bx)$ were pre-trained using \eqref{implicit}, where we chose the standard deviations \mrpar{R1.C6}{from a uniform distribution} ranging from $0$ to $0.1$. \mrpar{\\R4. C3}{ The i-MuSE algorithm and its accelerated version requires determining the Lipschitz  constant of $\mathcal{I}_\theta(\bx)$ to implement the update step \eqref{mm_imuse} for image recovery. Based on the reasoning given in the previous section, we determined the approximate value of the true Lipschitz constant using the CLIP approach which was found to be equal to $1.88$ and in the experiments we set $L=5$ which ensures convergence. Note that, the accelerated version of i-MuSE  also requires the Lipschitz constant $\beta$ of the mapping \eqref{proj}. Using the CLIP approach we found $\beta = 1.2$ and since the mapping removes noise from the iterate we used the Lipschitz constant $\beta$ given by CLIP.}
\subsubsection{Score network used in PnP models for comparison} We compare the proposed energy model with PnP-ISTA with the following update rule \cite{pnp_ista}:
\begin{eqnarray}\label{pnp_ista}
      Z_t = \bx_{t-1} - \gamma \dfrac{\bA^H(\bA\bx_{t-1}-\bb)}{\eta^2}\\\nonumber
      \bx_t = D_\sigma(Z_t)     
\end{eqnarray}
where $D_\sigma(\cdot): \mathbb C^m\rightarrow \mathbb C^m$ is a CNN-based denoiser. We  choose $D_\sigma(\cdot)$ as a five-layer CNN network with each layer having $64$ channels with $3\times 3$ filter size. Except for the output layer, between each layer ReLU activation function was used. The denoiser $D_\sigma(\cdot)$ was trained to remove zero-mean Gaussian noise  with standard deviation $\sigma=0.01$. In this work, we consider two kinds of denoiser: a) without contraction constraint (referred as PnP-ISTA), and b) with contraction constraint which was imposed using the spectral normalization technique. The latter is enforced to ensure the convergence of the PnP model. We refer to this algorithm as PnP-ISTA-SN. Furthermore, we also compared the performance of PnP model \mrpar{R4.C9}{when the five-layer CNN network was replaced with a deeper DRUNet (without contraction constraint) and was trained using \eqref{implicit} where $\nabla_\bx \mathcal{I}_{\theta}(\cdot)$ is replaced with $D_\sigma(\cdot)$  to remove Gaussian noise with standard deviations ranging from $0$ to $0.1$. This model is referred as PnP-ISTA (DRUNet). }

\subsubsection{E2E-MoDL} The energy models are also compared with MoDL, which is an unrolled algorithm and trained in an E2E fashion for $10$ iterations. A five-layer CNN was used to implement MoDL.

\subsection{Visualization of implicit and explicit energy functions}
\begin{figure}
\centering
\begin{subfigure}{0.48\linewidth}
    \includegraphics[width=1\linewidth]{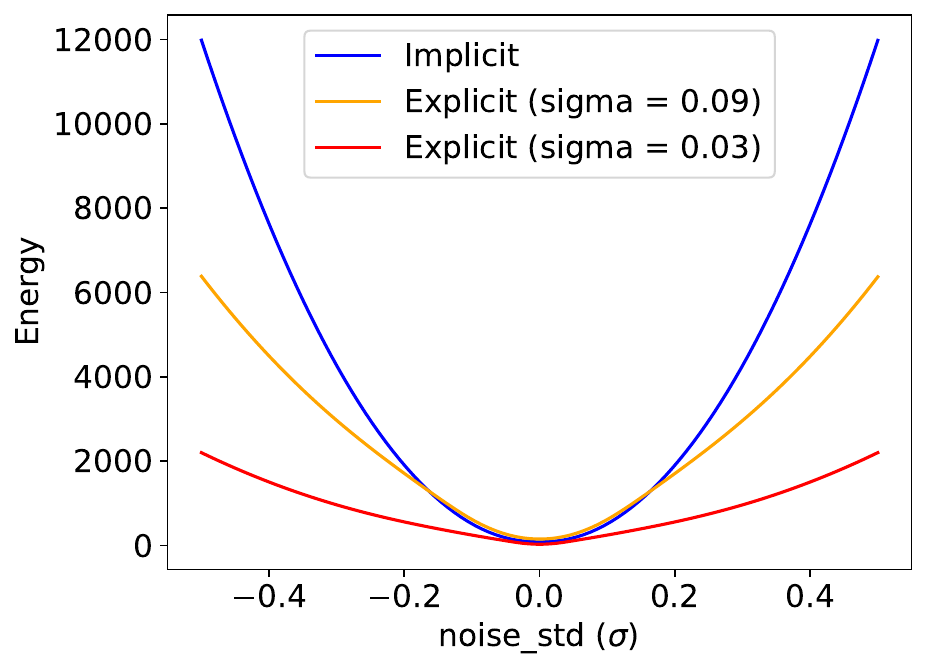}
    \caption{}
\end{subfigure}
\begin{subfigure}{0.48\linewidth}
    \includegraphics[width=1\linewidth]{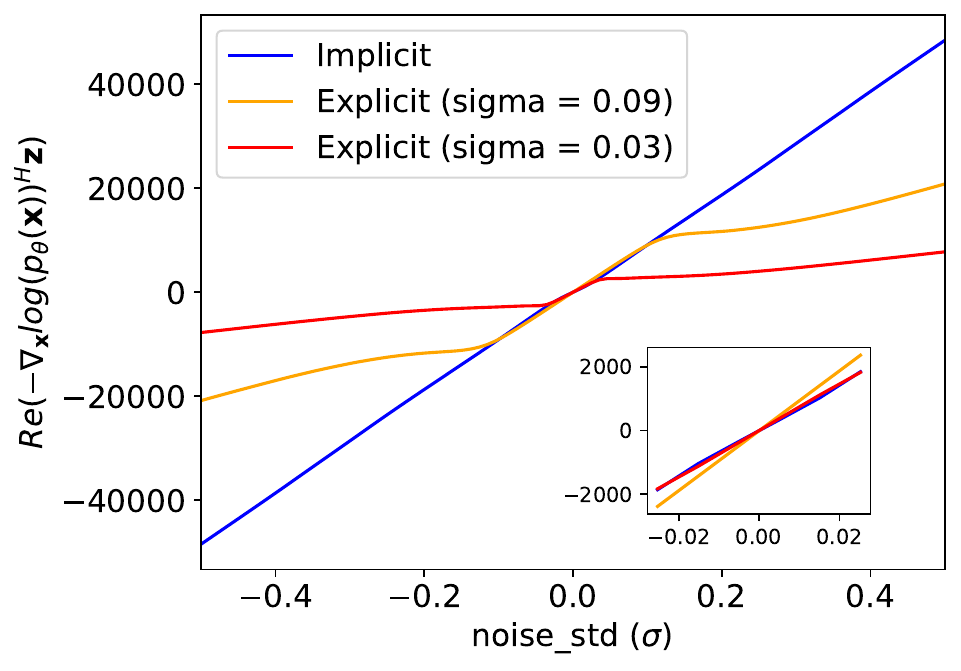}
    \caption{}
\end{subfigure}\\
\begin{subfigure}{0.48\linewidth}
    \includegraphics[width=1\linewidth]{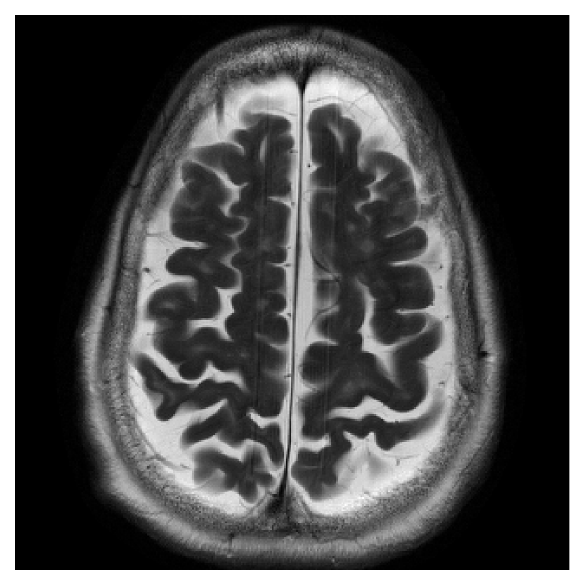}
    \mrpar{R4.M8}{\caption{}}
\end{subfigure}
\begin{subfigure}{0.48\linewidth}
    \includegraphics[width=1\linewidth]{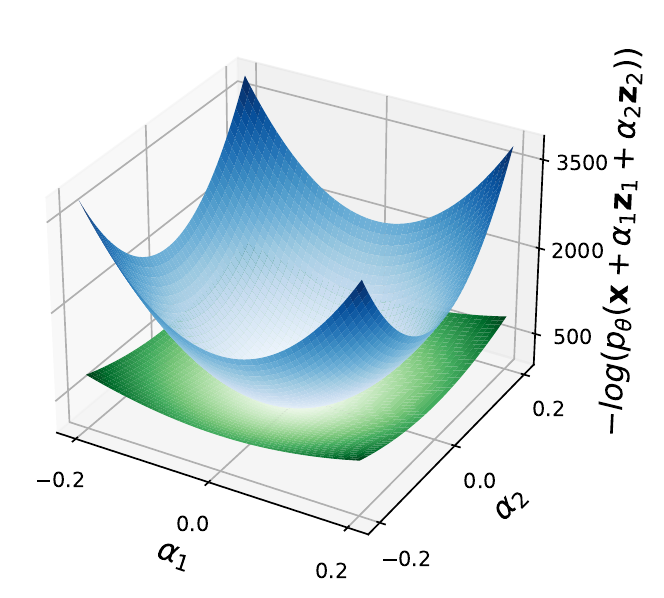}
    \caption{}
\end{subfigure}
\caption{(a) Implicit and explicit EM (for scales $0.09$ and $0.03$) vs. noise standard deviation $\sigma$, corresponding to noise added versions $\bx+\sigma \bz$ of the T2 weighted image $\bx$ in (c). (b) Plot of the inner product between the score of $\mathcal I_{\theta}$ and the Gaussian perturbations $\bz$,  compared with explicit EM for the scales $0.09$ and $0.03$, respectively.  (d) Multidimensional illustration of the implicit (blue surface) and explicit EM ($\sigma=0.03$ in green) energies, corresponding to the input $\tilde{\bx}=\bx+ \alpha_1 Z_1 + \alpha_2 Z_2$.  }
\label{Quadratic_illustration}
\end{figure}

In this section, we visualize the learned implicit and explicit multi-scale energy functions for scales $\sigma=0.03$ and $\sigma=0.09$. In particular, we feed the energy functions with $\tilde{\bx} = \bx +\sigma\bz$, where $\bx$ is a test image (whose maximum magnitude is one) shown in Fig.\ref{Quadratic_illustration}.c, $\bz\in \mathcal{N}(0,\bI)$ is a realization of zero-mean Gaussian noise, and $\sigma$ is the noise standard deviation that is varied from $-0.5$ to $0.5$. The plot of the energies as a function of $\sigma$ are shown in Fig.\ref{Quadratic_illustration}.a, while the corresponding score function is illustrated in Fig.\ref{Quadratic_illustration}.b. In particular, we plot the inner product between the derivative of the energy (score) with $\bz$. A multidimensional example is shown in Fig.\ref{Quadratic_illustration}.d, where the input is a perturbation of the form $\tilde{\bx}=\bx+ \alpha_1 Z_1 + \alpha_2 Z_2$, where $Z_1, Z_2 \sim \mathcal{N}(0,\bI)$ and $\alpha_1, \alpha_2$ are varied from $-0.2$ to $0.2$.

We observe from Fig.\ref{Quadratic_illustration}.a and Fig.\ref{Quadratic_illustration}.d that the i-MuSE energy is a quadratic function of $\sigma$ for all the $\sigma$ values. This validates Proposition \ref{lemma2}, which showed that the implicit energy function is  $\mathcal{I}_\theta(\tilde{\bx}) =\dfrac{1}{2} \sigma^2\|\bz\|^2$. In contrast, explicit energy functions only closely match the implicit function for the $\sigma$ values they were trained for, while they deviate for other standard deviations. The saturating nature of the explicit energy function with $\sigma$ can also be appreciated from Fig.\ref{Quadratic_illustration}.b, which shows that the gradient of i-MuSE is linear, while the gradient of e-MuSE energies are only linear close to the origin. The flatter nature of the e-MuSE energy away from the minimum translates to slower convergence, when initialized away from the true minimum. The zoomed sections in Fig. \ref{Quadratic_illustration}.(b) show that the e-MuSE gradient with $\sigma=0.09$ deviates from the line close to the origin, while i-MuSE and e-MuSE with $\sigma=0.03$ are on the straight line. 

\subsection{Study on convergence and sensitivity to initialization}
\mrpar{R3.C3}{In this section, we compare the convergence rate of single-scale EM, e-MuSE, i-MuSE, and i-MuSE (accelerated) (discussed in Section \ref{speedup}) for two different sampling rates. The single-scale, implicit, and each scale of the explicit algorithm were made to run until ${|f(\bx_{n+1})-f(\bx_n)|}/{|f(\bx_n)|} \leq 1e^{-7} $, where $f(\bx)$ in the case of single-scale EM, i-MuSE, i-MuSE (accelerated), and e-MuSE is the cost function defined in \eqref{eq:P1}, \eqref{imuse}, \eqref{imusemod}, and \eqref{eq:P2}, respectively.} For each setting, we determined the optimal $\eta$ and $\zeta$ for the algorithms based on their performance on a validation set. We consider two different flavors of initialization: 
\underline{(a). SENSE \cite{sense}:} We initialize the algorithm by $\bx_0= (\bA^H \bA +\tilde{\lambda}\bI)^{-1} (\bA^H \bb)$, which is the minimizer of the cost function $\dfrac{1}{2}\|
\bA\bx-\bb\|^2 +\tilde{\lambda}\|\bx\|^2$. Since the SENSE solution satisfies the DC term, it is closer to the true solution.
\underline{(b). Random initialization:} Three different random initializations were obtained from a zero-mean Gaussian distribution. In this case the initialization is far from the true solution. 

\begin{figure}
\centering
\begin{subfigure}{0.48\linewidth}
    \includegraphics[width=1\linewidth]{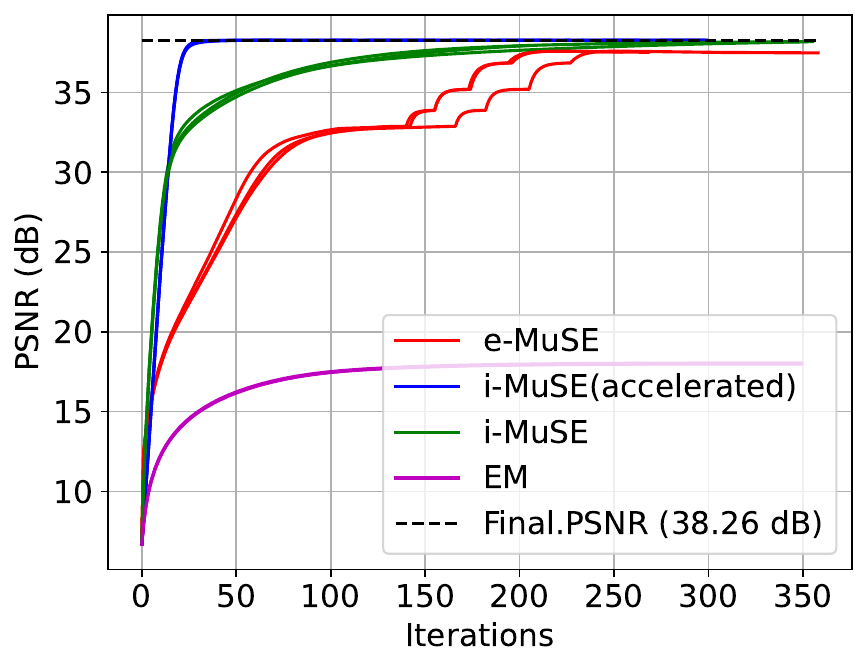}
    \caption{4x: random initialization}
\end{subfigure}
\begin{subfigure}{0.48\linewidth}
    \includegraphics[width=1\linewidth]{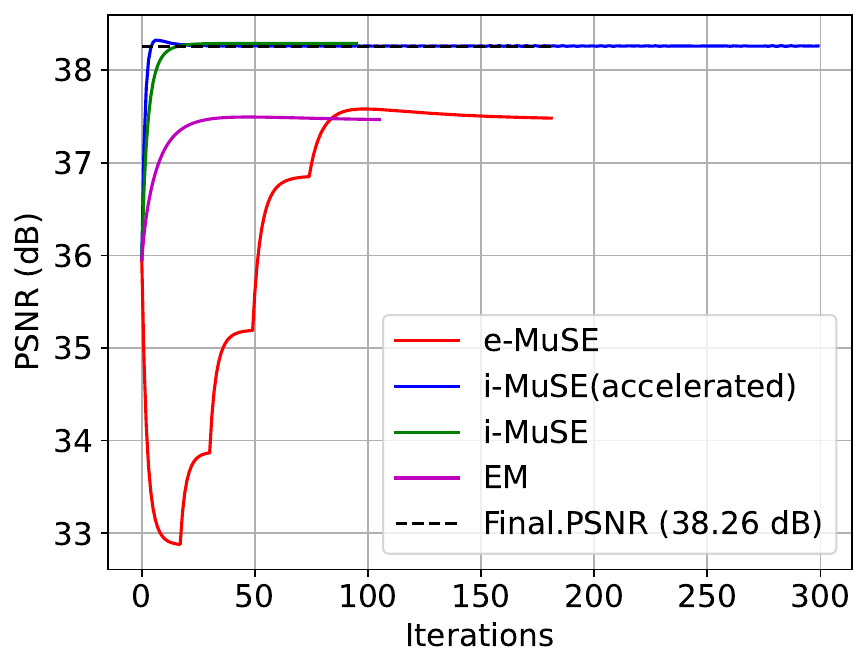}
    \caption{4x: SENSE initialization}
\end{subfigure}
\begin{subfigure}{0.48\linewidth}
    \includegraphics[width=1\linewidth]{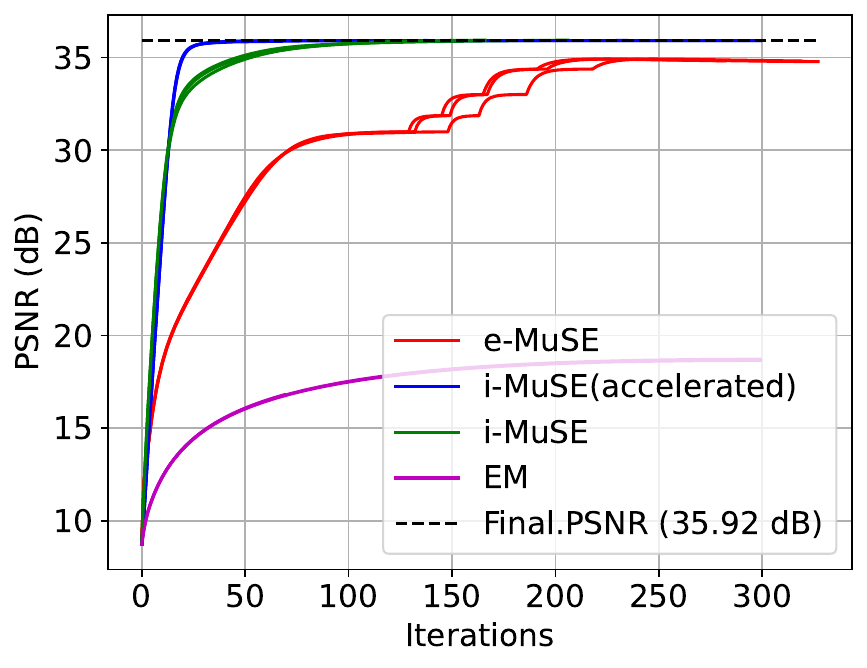}
    \caption{6x: random initialization}
\end{subfigure}
\begin{subfigure}{0.48\linewidth}
    \includegraphics[width=1\linewidth]{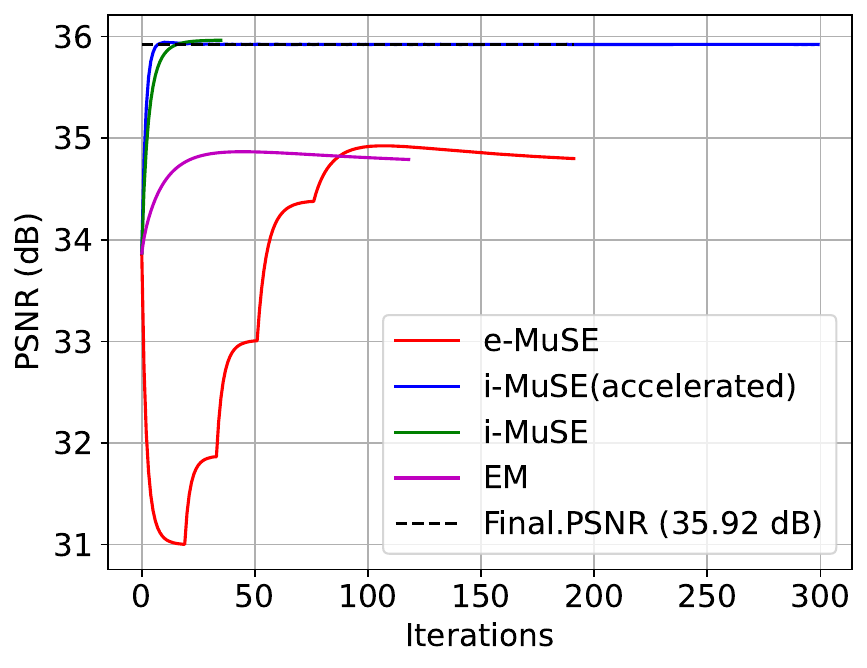}
    \caption{6x: SENSE initialization}
\end{subfigure}
\caption{Convergence plot of the multi-scale algorithms for (a) 4-fold acceleration and three different random initializations, (b) 4-fold acceleration and SENSE initialization, (c) 6-fold acceleration and three different random initializations, and (d) 6-fold acceleration and SENSE initialization. The blue, green, magenta, and red lines denote the PSNR evolution of i-MuSE (accelerated), i-MuSE, single-scale EM, and e-MuSE, respectively. In the legend "Final PSNR" denotes the final PSNR value of i-MuSE algorithm.} 
\label{convergence}
\end{figure}

Fig. \ref{convergence} compares the convergence rate of the single-scale EM, explicit and implicit algorithms. We observe that i-MuSE (accelerated) converges faster than the other single-scale and multi-scale algorithms, regardless of initialization. One may resort to hyperparameter optimization of the convergence thresholds at different scales to improve the convergence rate of e-MuSE, which is not required for i-MuSE. We note that when SENSE initialization is used, the i-MuSE (accelerated) scheme converges rapidly (in around 20 iterations)\footnote{\mrpar{R4.C3}{We note that the convergence rate depends on the Lipschitz constant $L$. Estimating the true Lipschitz constant could result in faster convergence of each scale of e-MuSE and i-MuSE. However, this is beyond the scope of the current work.}}. On the contrary, the e-MuSE scheme takes significantly longer and converges to a solution with a lower Peak Signal-to-Noise Ratio (PSNR). In particular, the optimization at the coarser scales moved the SENSE solution away from the minimum. Although the performance of e-MuSE may be improved in this case by carefully choosing the specific scales used in the algorithm, the i-MuSE scheme does not need this careful optimization. We also observe that the performance of single-scale EM dropped significantly when initialized randomly, while the performance of multi-scale algorithms remains almost unaffected by the initialization. Finally, note that i-MuSE with and without acceleration converges to similar PSNR values. Therefore, in the remaining experiments we consider the accelerated version of i-MuSE.

\subsection{Comparison with SOTA MAP approaches}
In this section, we compare the reconstruction performance of the proposed energy models with E2E trained MoDL and PnP-ISTA.  The parameters $\eta$, $\zeta$ of e-MuSE and i-MuSE algorithms were kept the same as in the previous experiment. The energy-based algorithms were run until their respective cost functions satisfied  ${|f(\bx_{n+1})-f(\bx_n)|}/{|f(\bx_n)|} \leq 1e^{-5} $ or until $500$ iterations were reached. Similarly, PnP-ISTA was run until ${|\bx_{n+1}-\bx_n|}/{|\bx_n|} \leq 1e^{-5} $ or until $500$ iterations \mrpar{R4.M7}{were} reached.  

We evaluated the reconstruction algorithms on $4$ and $6$-fold acceleration for two different initializations: SENSE and random samples drawn from zero-mean Gaussian distribution. Table \ref{recon_cmp} compares the algorithms using PSNR and Structural Similarity Index Measure (SSIM) metrics. 

We observe that i-MuSE performs consistently well across all the settings, with performance comparable to the E2E-trained MoDL. However, it is worth pointing out that i-MuSE learns a generic prior while MoDL has to be trained separately for each setting. The sensitivity of MoDL to the forward operator is illustrated in Section \ref{cartesian}. 

%We observe that the E2E-MoDL, which involved the end-to-end training of the network, offered the best PSNR and SSIM across all settings, compared to the other methods. However, we note that all the other models learned a generic prior, unlike MoDL needs to be trained for each acquisition setting separately. Nevertheless, we note that performance of the i-MuSE algorithm is only marginally worse than E2E-MoDL in all the settings.

\mrpar{R4.C9}{We notice that with SENSE initialization, the performance of i-MuSE, PnP-ISTA, PnP-ISTA (DRUNet), and E2E-MoDL are roughly comparable in the 4x accelerated setting. By contrast, the spectral normalization used in PnP-ISTA-SN network translates to reduced performance. We observe that the performance of PnP-ISTA and PnP-ISTA (DRUNet) drops significantly in the 6x setting, compared to i-MuSE and MoDL. We note that there are visible artifacts in the PnP-ISTA reconstructions in the 6x setting. See Fig. \ref{acc_cmp}.b where PnP-ISTA has a bright spot in the reconstructions, compared to PnP-ISTA-SN. We attribute these artifacts to the lack of contraction constraints on the networks, which are required for the guaranteed convergence of PnP algorithms. We observe that PnP-ISTA (DRUNet) performance is worst in the 6x setting, even with  SENSE initialization. This is because the network is considerably deeper than that of a five-layer CNN, which may translate to a high Lipschitz constant, resulting in divergence. We also note that the performance of single-scale approaches (EM and PnP-ISTA) falls significantly when initialized randomly. In particular, the energy may have zero or low gradients closer to the initialization, which restricts the convergence to the minimum. By contrast, the performance of the multi-scale approaches (e-MuSE and i-MuSE) remained largely unaffected at moderate accelerations, indicating that they converged to a unique minimum of the posterior energy.}

Fig. \ref{acc_cmp}.a and Fig. \ref{acc_cmp}.b show the reconstructed images for accelerations $4$-fold and $6$-fold with SENSE initialization, respectively. We note that all models offer good performance in the 4x setting, while single-scale methods such as EM, PnP-ISTA, PnP-ISTA-SN exhibit higher errors. We note that the PnP-ISTA reconstruction shows a bright spot, which translated into a lower PSNR.

\subsection{Generalization performance}\label{cartesian}
\begin{figure*}[!h]
    \centering
    \begin{subfigure}[b]{0.8\textwidth}  
        \includegraphics[width=1\linewidth]{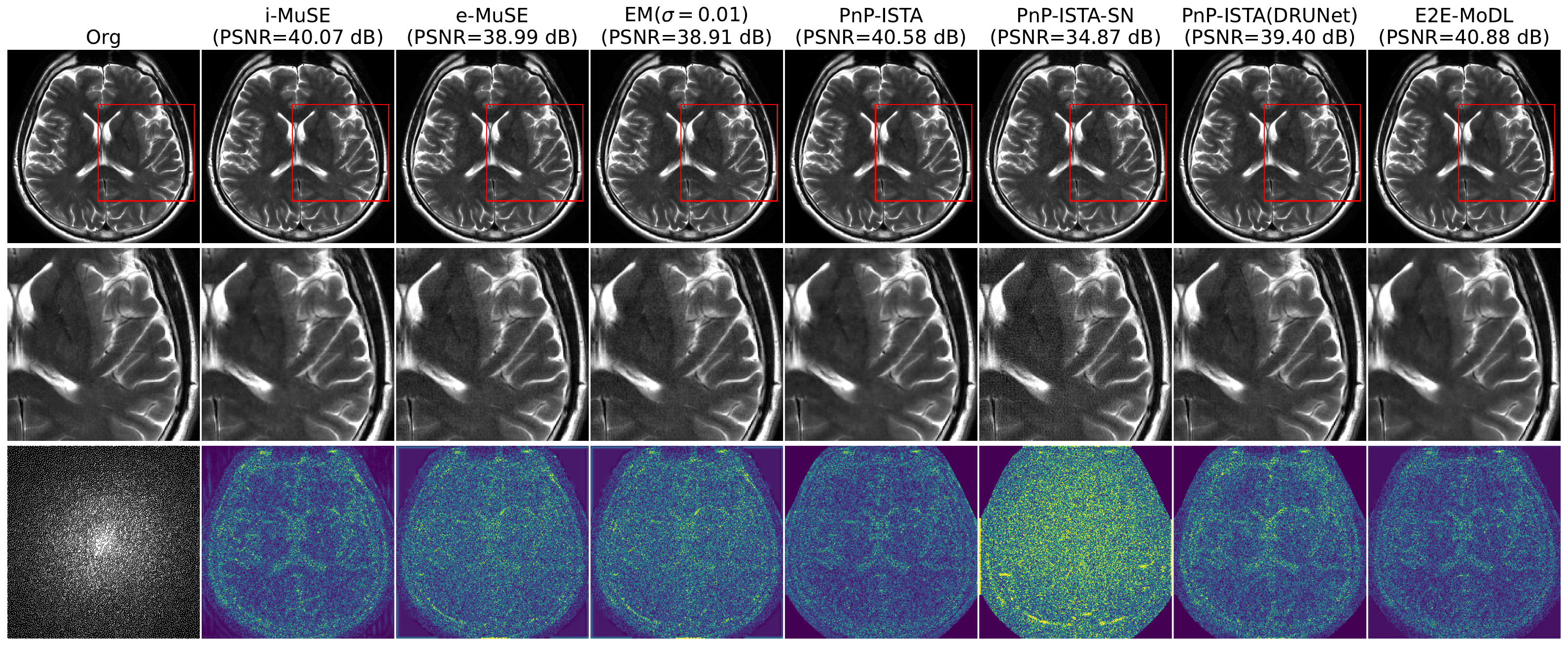}
    \caption{Four-fold acceleration}
    \label{two_fold}
    \end{subfigure}
       \begin{subfigure}[b]{0.8\textwidth} 
        \includegraphics[width=1\linewidth]{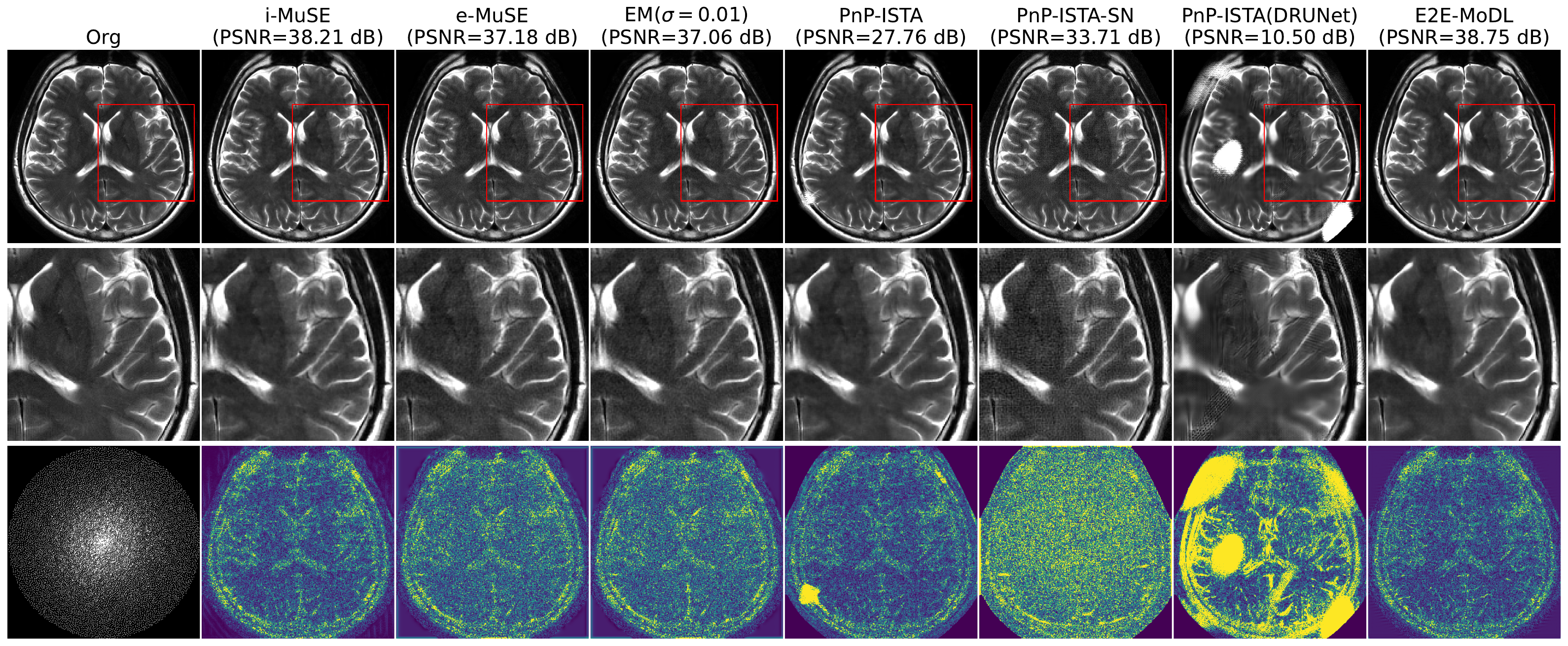}
    \caption{Six-fold acceleration}
    \label{four_fold}
    \end{subfigure} 
    \caption{Comparison of multi-scale energy models i-MuSE, e-MuSE with the single-scale energy model EM, PnP-ISTA (denoiser trained using a five-layer CNN), PnP-ISTA-SN (denoiser trained using a five-layer CNN with contraction constraint), PnP-ISTA (DRUNet) (denoiser trained using DRUNet), and E2E trained MoDL at two different accelerations on the fastMRI brain data set. MoDL was trained separately for each acceleration, while we used the same energy and score networks for both accelerations. The top and second rows of each plot show the reconstructed and enlarged image, respectively. The first image in the last row of each plot is the undersampling mask, while the remaining images show the error images that are scaled by a factor of 10 to highlight the differences} \label{acc_cmp}.
\end{figure*}

\begin{figure}
\centering
    \includegraphics[width=0.4\textwidth]{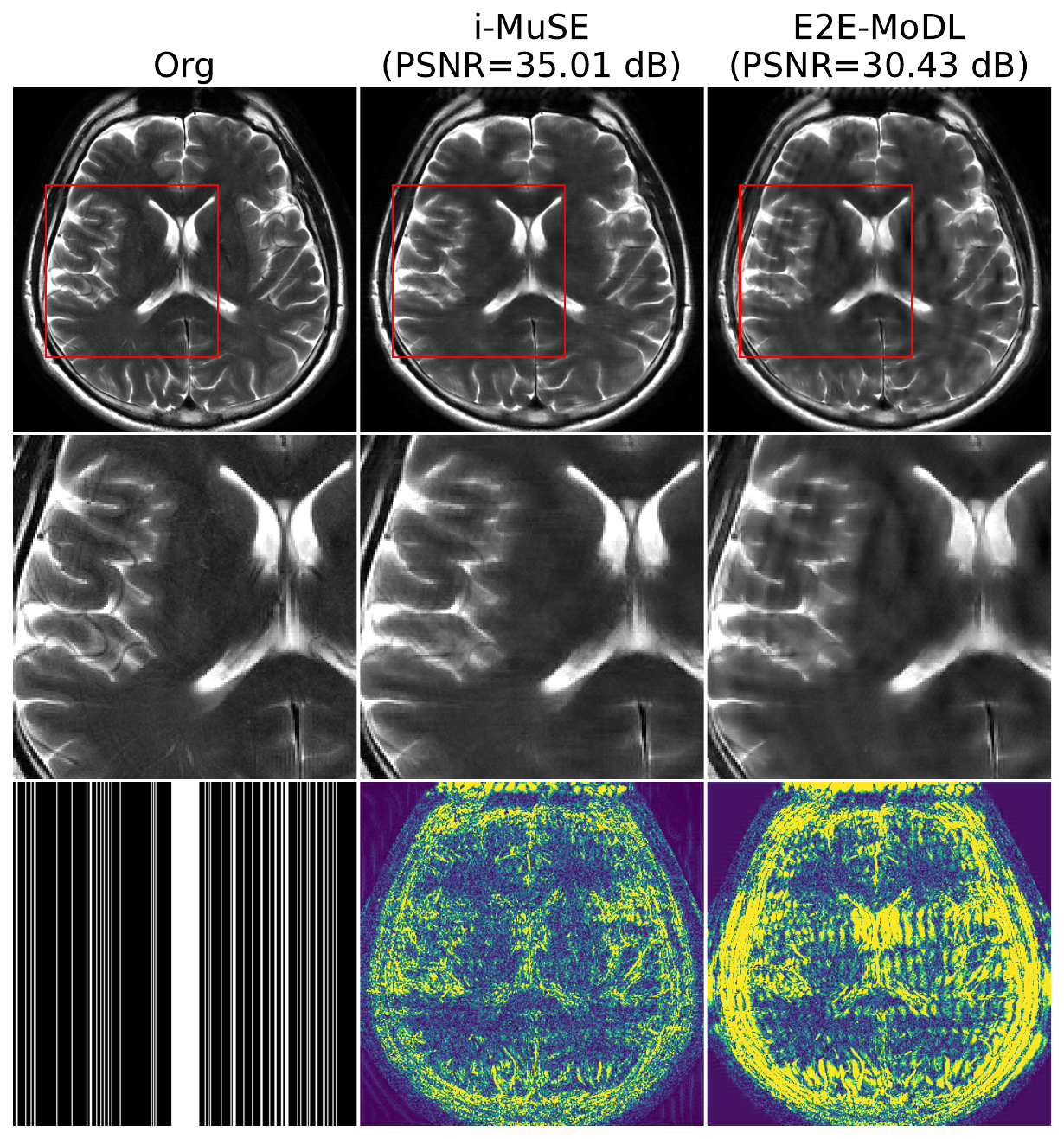}
    \caption{Illustration of generalization ability of energy model. MoDL was trained on 4-fold Poisson mask and tested on 4-fold Cartesian mask. Top and second row shows the reconstructed and enlarged image, respectively. The first image in the last row shows the Cartesian mask employed for undersampling, while the remaining images shows the error images which are scaled by a factor of $10$.}
    \label{Cartesian sampling}
\end{figure}

We now compare the generalization performance of E2E-MoDL with the i-MuSE algorithm. We trained MoDL using a 4-fold Poisson mask, while inference was performed on a \mrpar{R4.C7}{4-fold 1-D undersampling mask}. The results are shown in Table \ref{Flexib} and in Fig. \ref{Cartesian sampling}. We observe that unlike the case in Table \ref{recon_cmp} and Fig. \ref{acc_cmp}, where the training and inference settings are similar, the MoDL algorithm underperformed i-MuSE by around 4 dB. We also note from Fig. \ref{Cartesian sampling} that the MoDL reconstruction suffers from alias artifacts, \mrpar{R4.C6}{while i-MuSE is relatively insensitive to them in the experiments we have considered. We note that the relative insensitivity of algorithms to the forward models and initialization are desirable characteristics during clinical deployment.}
\begin{table*}[]
\centering
\caption{Comparison of the models in the context of accelerated MRI at two different accelerations for different initialization.}
\resizebox{\textwidth}{!}{%
\begin{tabular}{|lllll|llll|}
\hline
\multicolumn{5}{|c|}{\textbf{4x acceleration}}                                                                                                                                                  & \multicolumn{4}{c|}{\textbf{6x acceleration}}                                                  \\ \hline
\multicolumn{1}{|l|}{\multirow{2}{*}{Algorithm}} & \multicolumn{2}{c|}{SENSE initialization}                             & \multicolumn{2}{c|}{Random initialization}                           & \multicolumn{2}{c|}{SENSE initialization}         & \multicolumn{2}{c|}{Random initialization} \\ \cline{2-9} 
\multicolumn{1}{|l|}{}                           & \multicolumn{1}{c}{Avg. PSNR $\pm$ std (dB)} & \multicolumn{1}{c|}{SSIM}  & \multicolumn{1}{c}{Avg. PSNR $\pm$ std (dB)} & \multicolumn{1}{c|}{SSIM} & Avg. PSNR + std (dB) & \multicolumn{1}{l|}{SSIM}  & Avg. PSNR +std (dB)         & SSIM         \\ \hline
\multicolumn{1}{|l|}{i-MuSE}                    &                                       $39.36 \pm 1.34$   & \multicolumn{1}{l|}{0.98} & $39.19 \pm 1.36$                                         & 0.979                     & $37.63 \pm 1.25$                     & \multicolumn{1}{l|}{0.978} & $37.50 \pm 1.23$                            & 0.975      \\ 
\multicolumn{1}{|l|}{e-MuSE}                     &       $38.43 \pm 1.47$                                   &  \multicolumn{1}{l|}{0.97}      &   $38.38 \pm 1.48$                                       &   0.972                        & $36.73 \pm 1.37$                     & \multicolumn{1}{l|}{0.967}      & $36.54 \pm 1.36$                            &  0.964            \\ 
\multicolumn{1}{|l|}{EM}                         &$38.44 \pm 1.48$                                         & \multicolumn{1}{l|}{0.97}      & $17.67 \pm 2.02$                                          &  0.73                        &     $36.64 \pm 1.35$                 & \multicolumn{1}{l|}{0.967}      &  $17.58 \pm 2.00$                           &  0.725            \\ 
\multicolumn{1}{|l|}{PnP-ISTA}                   &              $39.89 \pm 1.38$                            & \multicolumn{1}{l|}{0.97}      &  $-18.31\pm2.2$                                    &    0.66                       & $36.78 \pm 5.13$                     & \multicolumn{1}{l|}{0.96}      & $-24.31 \pm 2.06$                           &   $-0.0001$           \\
\multicolumn{1}{|l|}{PnP-ISTA-SN} &              $34.94 \pm 1.34$                            & \multicolumn{1}{l|}{0.94}      & $25.87\pm1.69$                                        &    0.75                       & $34.11 \pm 1.25$                     & \multicolumn{1}{l|}{0.93}      & $16.95 \pm 1.44$                            &   $0.44$           \\
\multicolumn{1}{|l|}{\color{black}PnP-ISTA(DRUNet)} &              \color{black}$38.59 \pm 1.22$                            & \multicolumn{1}{l|}{\color{black} 0.97}      &  \color{black}$38.59\pm1.22$                                        &   \color{black} 0.97                       & \color{black} $6.16 \pm 3.34$                     & \multicolumn{1}{l|}{0.8}      &\color{black} $6.09 \pm 3.37$                            &   \color{black} $0.8$           \\
\multicolumn{1}{|l|}{E2E-MoDL}                       &             $40.13 \pm 1.415$                             & \multicolumn{1}{l|}{0.98}      &     -                                     &      -                     & $38.3 \pm 1.2$                     & \multicolumn{1}{l|}{0.977}      &        -                     &           -   \\ \hline
\end{tabular}%
}
\label{recon_cmp}
\end{table*}

\begin{figure*}[!h]
    \centering
    \begin{subfigure}[b]{0.8\textwidth}  
        \includegraphics[width=1\linewidth]{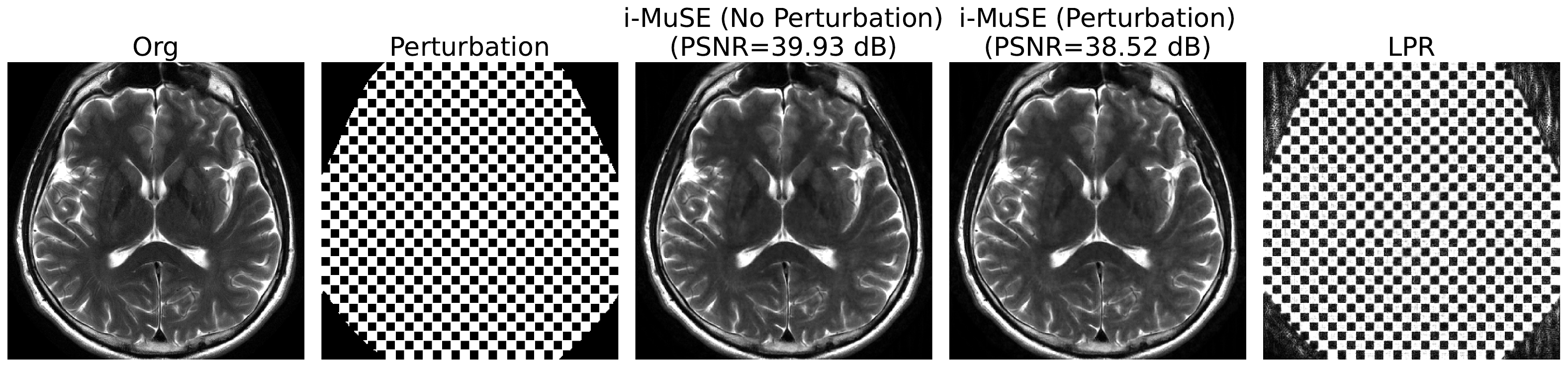}
    \caption{Four-fold acceleration}
    \label{four_fold_pert}
    \end{subfigure}
       \begin{subfigure}[b]{0.8\textwidth} 
        \includegraphics[width=1\linewidth]{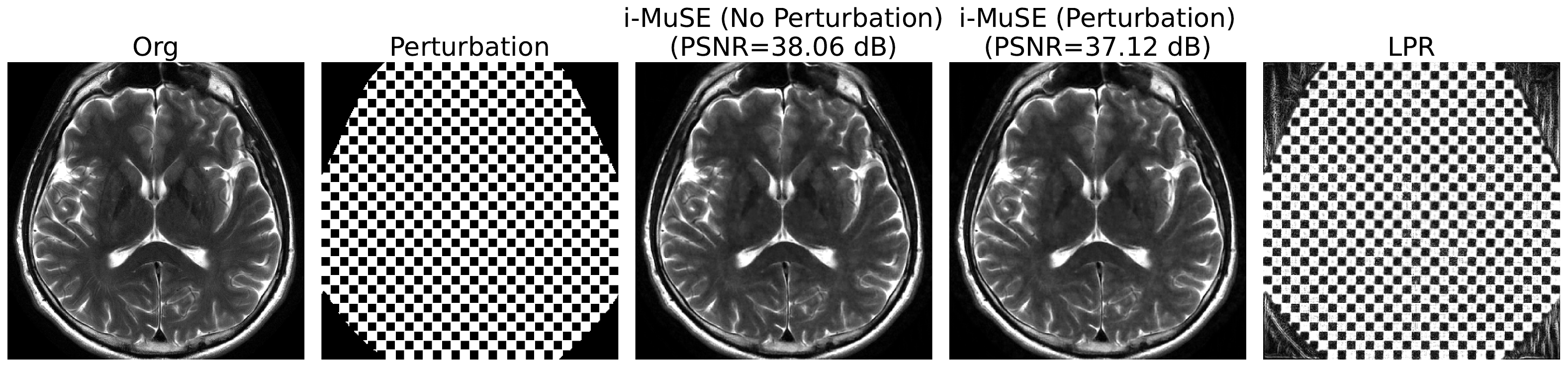}
    \caption{Six-fold acceleration}
    \label{six_fold_pert}
    \end{subfigure} 
    \caption{\textcolor{black}{Illustration of i-MuSE being robust to perturbation for (a) a four-fold acceleration and (b) a six-fold acceleration. Each row shows the reconstructed image with and without the added perturbation. The last column shows the LPR that looks similar to the added perturbation. This indicates that the proposed algorithm is robust to perturbation. }} \label{pert}.
\end{figure*}

\begin{table}[H]
\centering
\caption{Generalization performance of MuSE}
\begin{tabular}{|p{2cm}|p{3cm}|p{2cm}|}
\hline
Algorithm & Avg. PSNR $\pm$ std (in dB)&SSIM\\
\hline
i-MuSE & $33.35 \pm 1.69$&  0.956\\
% PnP-ISTA& $29.06 \pm 10.73$ SSIM=0.936\\ 
% \hline
E2E-MoDL & $29.89 \pm 1.51$& 0.928\\\hline
\end{tabular}
\label{Flexib}
\vspace{-1em}
\end{table}
\mrpar{R4.C2}{\subsection{Local Perturbation Response}
In this section, we evaluate the sensitivity of the i-MuSE algorithm to perturbations during inference. In particular, we added checkerboard-type perturbation $\bp$ to the reference image \cite{lpr}:
\begin{equation}
    \bb_\bp = \bA(\bx+\bp)+\bn = \bb + \bA(\bp)
\end{equation}
where $\bb_p$ represents the perturbed measurement. We simulated a checkerboard-type perturbation with magnitude $1e^{-2}$ and rectangle size $= 10$ voxels. Fig.\ref{pert}.a and Fig.\ref{pert}.b show the reconstructed images (with and without the perturbation) for a four-fold and a six-fold acceleration, respectively. In addition, Fig. \ref{pert} also shows the checkerboard-type perturbation and the Local Perturbation Response (LPR) obtained by evaluating the difference between the image reconstruction from the perturbation and the perturbation-free measurements. From the figure, it can be observed that the LPR looks similar to the added perturbation for both the accelerations. These results show that the MAP estimates of i-MuSE may not filter out important content present in the measurements, which will be problematic in a medical imaging setting.}
\section{Experiments: Sampling}\label{sampling_exps}
\subsection{Data set}
In this section, we determine the utility of multi-scale strategy in sampling from the prior and posterior distributions. We illustrate the performance of sampling algorithms on the {Fashion MNIST} \cite{fmnist} and the T2-weighted brain data set. The former data set consists of $28 \times 28$ real gray scale images and each image is associated with classes: shirt / top, trousers, pullovers, dresses, coats, sandals, shirts, sneakers, bags and ankle boots. In this work, we resized the real grayscale images to $32 \times 32$ and trained the sampling algorithms on $10,000$ representative images from the Fashion MNIST data set. We also evaluated the sampling algorithms in the context of inpainting with different masks applied on the reference image. 
\subsection{Architecture and Implementation}
\subsubsection{i-MuSE}
In the case of the Fashion MNIST data set, similar to \cite{msdm}, we used 12-layer ResNet with 64 filters to represent $\bvarphi_\theta(\bx)$ whose optimal parameters were trained using \eqref{implicit} to predict noise for geometrically distributed standard deviations in the range $[0.1, 3]$. In the case of T2-weighted brain data set we used the same energy network as in the previous section. We used the Langevin algorithm to sample the prior and posterior distributions, where the former case $\nabla_\bx \log p_\theta(\bx|\bb)$ in \eqref{Langevin} is replaced with $\nabla_\bx \log p_\theta(\bx)=-\mathcal{I}_\theta(\bx)$. The step size $\epsilon$ and the annealing strategy for the temperature $t$ in the Fashion MNIST data set were chosen the same way as suggested in \cite{msdm}. For the brain data set, the step size was set as $0.02$, and the temperature $t$ was annealed where we began with $t=1$, and after every $50$-th iteration, we updated $t=\max(1e^{-4},0.2t)$. The temperature and step size were found empirically. We also studied the sampling performance when an unconditional score network was trained using \eqref{implicit} wherein $\nabla_\bx \mathcal{I}_\theta(\bx)$ is replaced by the 12-layer ResNet.
\subsubsection{Diffusion models}
\mrpar{R1.C4, R5.C2}{We compared the  i-MuSE sampling performance with the diffusion model \cite{yansong} that samples from the prior and the Diffusion Posterior Sampling (DPS) \cite{dps}, which, as the name suggests, samples from the posterior distribution}. The time-conditional score network for both methods is chosen as a 12-layer ResNet. During training, noise was added to the data according to the following perturbation kernel:
\begin {equation}\label{pert_score}
p_{0t}(\bx(t)|\bx(0))=\mathcal{N}(\bx(t);\bx(0),\dfrac{1}{2\log\Lambda}(\Lambda^{2t}-1)\bI)
\end{equation}
where $t$ is uniformly sampled over $[0,T]$, $\{\bx(t)\}_{t=0}^{T}$ represents the diffusion  process, $p_{0t}(\bx(t)|\bx(0))$ represents the transition probability from $\bx(0)$ to $\bx({t})$, and $\bx(0)$ are the samples from the training data set. In this work, we set $T=1$ and $\Lambda=25$ as suggested in \cite{yansong} The optimal parameters of the time-conditional score network $s_\theta(\bx(t),t )$ are learned by using the loss proposed in \cite{yansong}. We also investigate the performance of sampling when a 12-layer ResNet (without time conditioning) is trained as a score network using the loss proposed in \cite{yansong}. In case of sampling from the prior distribution, we implemented the Euler-Maruyama solver \cite{yansong} . To sample from the posterior distribution, we used the sampling algorithm proposed in \cite{dps}, where an additional step of update of the data guidance is added. 
\subsection{Sampling from the prior distribution}
\begin{figure}[t!]
  \centering
  \begin{subfigure}[b]{0.45\linewidth}
    \includegraphics[width=1\linewidth]{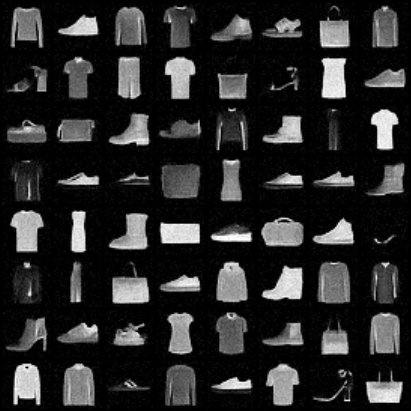}
    \caption{i-MuSE}
  \end{subfigure}
  \begin{subfigure}[b]{0.45\linewidth}
    \includegraphics[width=1\linewidth]{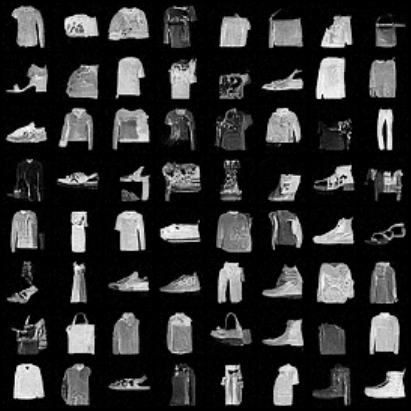}
    \caption{Cond. score \cite{yansong}}
  \end{subfigure}
  \begin{subfigure}[b]{0.45\linewidth}
    \includegraphics[width=1\linewidth]{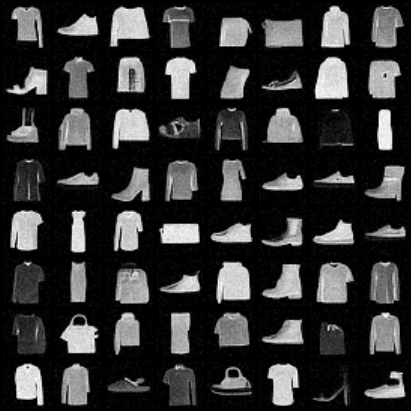}
    \caption{Uncond. score using \eqref{implicit}}
  \end{subfigure}
  \begin{subfigure}[b]{0.45\linewidth}
    \includegraphics[width=1\linewidth]{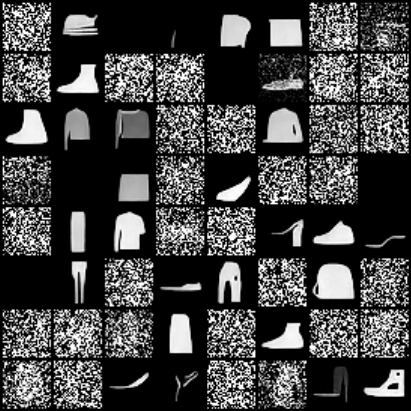}
    \caption{Uncond. score using \cite{yansong}}
  \end{subfigure}
  \caption{Fashion MNIST samples generated using (a) i-MuSE (b) diffusion model which employs time-conditional score network (c) unconditional score network trained using the loss in \eqref{implicit}, and (d) unconditional score network trained using the loss proposed in \cite{yansong}.}
\label{sampling_prior}
\end{figure}

\begin{figure}[htbp]
    \centering
    \begin{subfigure}[a]{0.45\textwidth}
        \centering
        \includegraphics[width=1\textwidth]{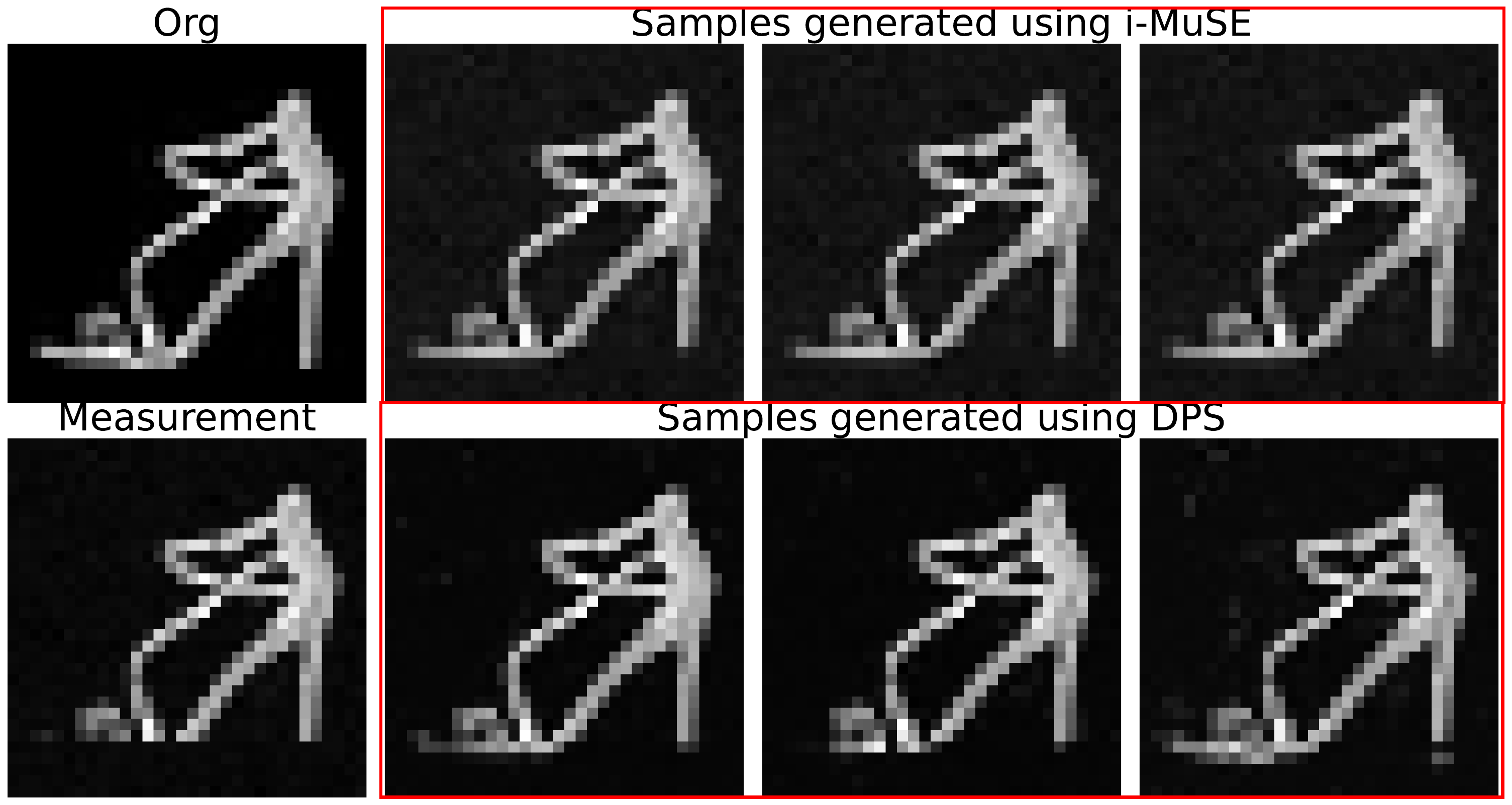}
        \caption{}
        \label{fig:sub1}
    \end{subfigure}
    \hspace{-1.5mm}
    \begin{subfigure}[b]{0.45\textwidth}
        \centering
        \includegraphics[width=1\linewidth]{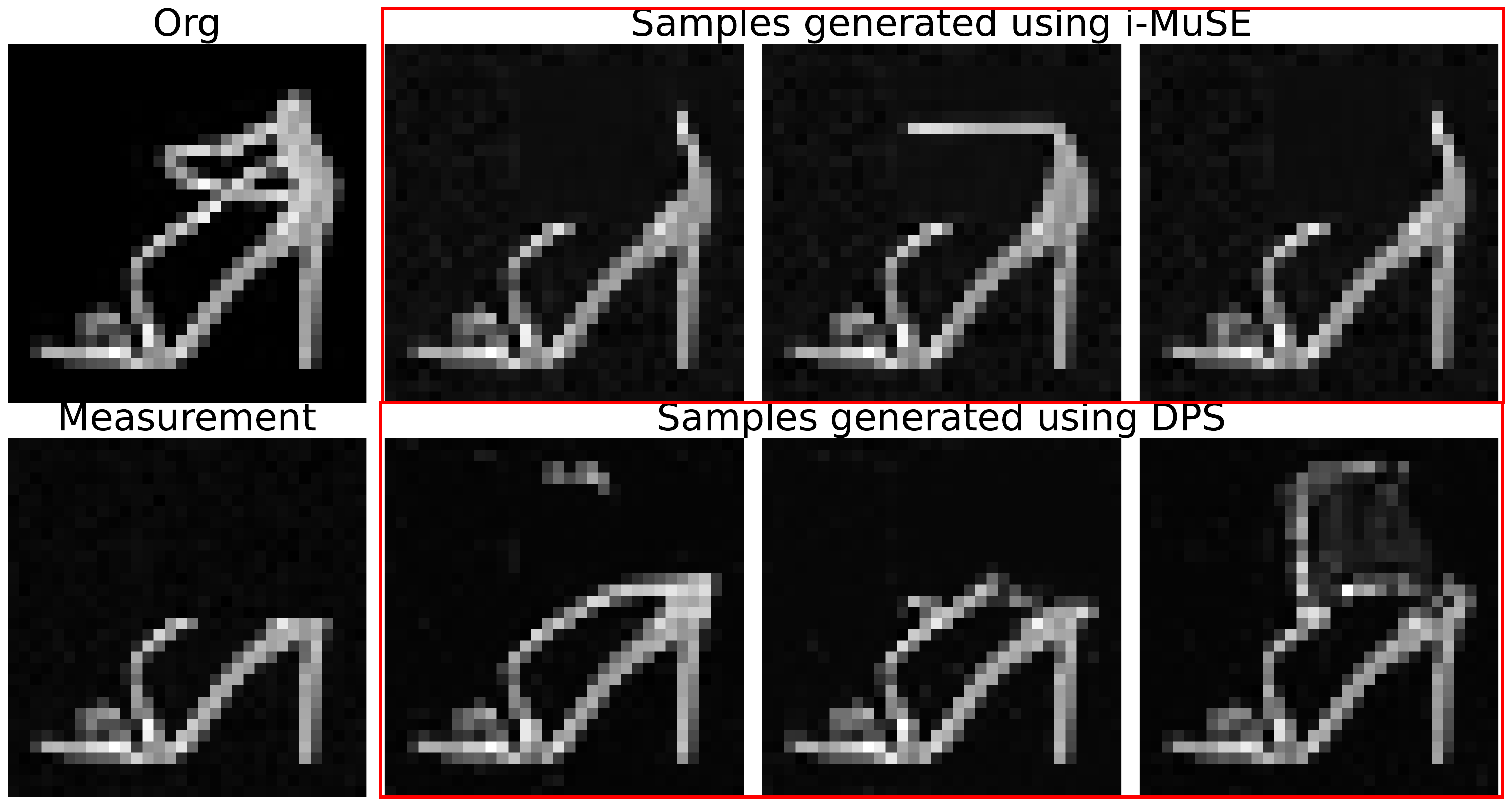}
        \caption{}
        \label{fig:sub2}
    \end{subfigure}
    \begin{subfigure}[c]{0.45\textwidth}
        \centering
        \includegraphics[width=1\linewidth]{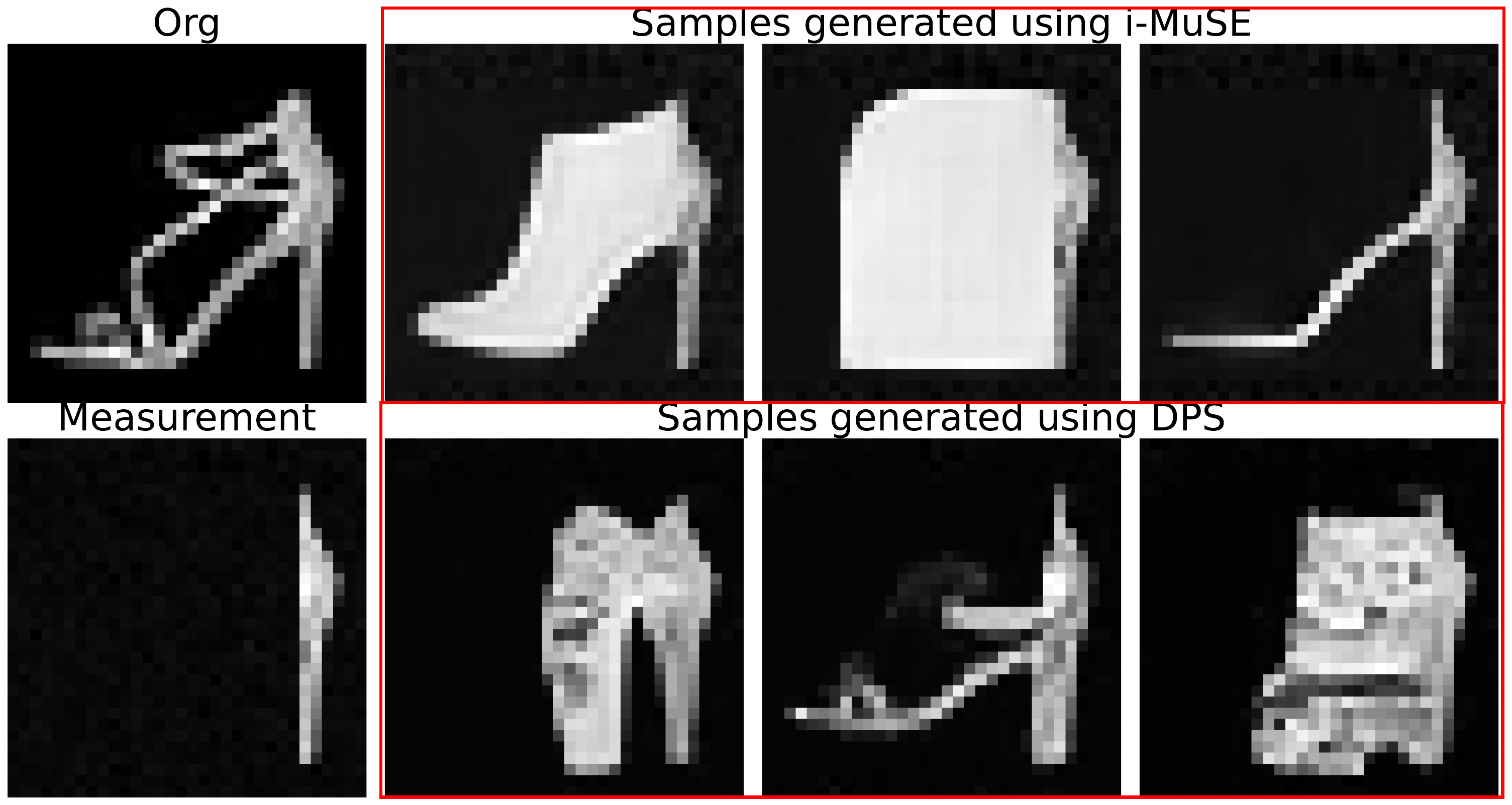}
        \caption{}
        \label{fig:sub3}
    \end{subfigure}
    \caption{Illustration of sampling from the posterior distribution in the context of inpainting on the Fashion MNIST data set for different masks. The first column in each figure shows the original image and the corresponding measurement, obtained by masking the respective regions. The remaining images (outlined by red boxes) show the samples generated by i-MuSE (top row) and DPS (bottom row), respectively. The size of the mask increases from (a) to (c), corresponding to increasing difficulty in recovery. We note that all the generated images are consistent with the measurements, while qualitatively similar to the Fashion MNIST images.}
    \label{sampling_inpainting}
\end{figure}

\begin{figure*}[!h]
    \centering
    \begin{subfigure}[b]{0.75\textwidth}  \includegraphics[width=1\linewidth]{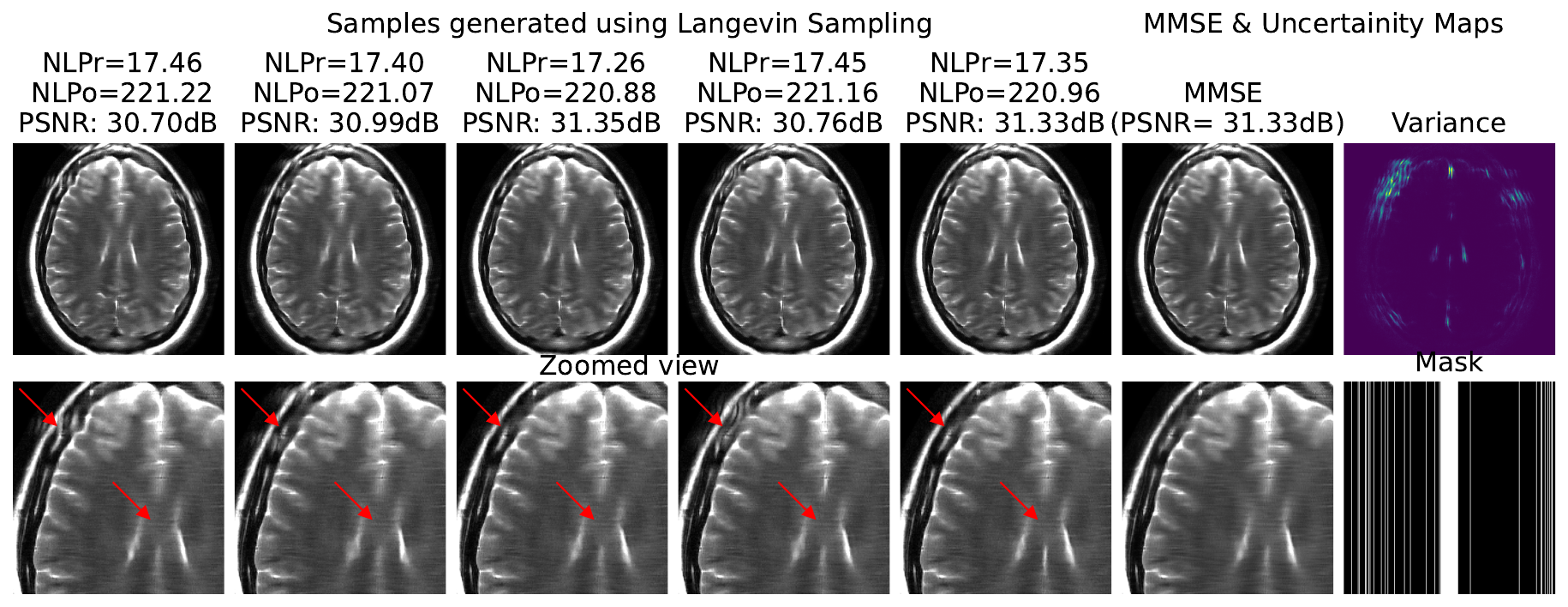}
    \caption{Six-fold acceleration}
    \label{two_fold}
    \end{subfigure}
       \begin{subfigure}[b]{0.75\textwidth} \includegraphics[width=1\linewidth]{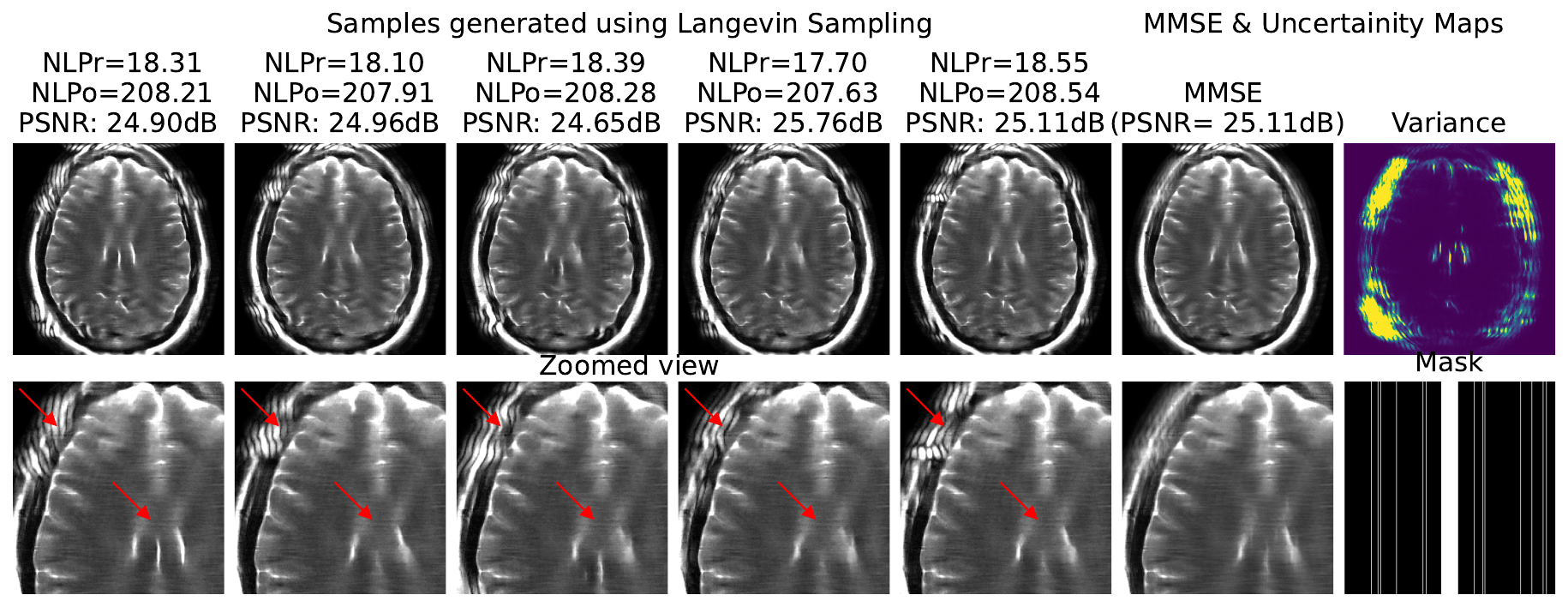}
    \caption{Eight-fold acceleration}
    \label{four_fold}
    \end{subfigure} 
    \caption{: Illustration of sampling from the posterior distribution in the context of MR image reconstruction on the T2-weighted brain data set when the image is undersampled using 6x and 8x 1-D Cartesian undersampling pattern. We chose aggressive acceleration to demonstrate the variability, while lower acceleration (not shown in the paper) showed very low variance. The first rows of each graph show the samples generated via the Langevin algorithm. The second row show the zoomed view of the corresponding sample. The last two columns of the first row show the MMSE and the variance map. We also show the sampling mask below the variance map.}
    \label{sampling}
\end{figure*}

\mrpar{R2.C0\\R3.C2}{In this section, we compare the sampling performance of i-MuSE with the diffusion model proposed in \cite{yansong}, and two variants of the unconditional score network whose optimal parameters are learned using \textbf{(a)} \eqref{implicit}, where the score predicts $\sigma \bz$ and \textbf{ (b)} the loss proposed in \cite{yansong}, where the score predicts ${\bz}/{\sigma}$. The sampling algorithms were initialized randomly from the Gaussian distribution for all models. Fig. \ref{sampling_prior}.a - Fig. \ref{sampling_prior}.d shows the Fashion MNIST samples generated using i-MuSE, diffusion model, and the unconditional score models trained using \eqref{implicit} and \cite{yansong}, respectively.  We observe that all models (except the unconditional score model using the traditional score matching loss in \cite{yansong}) can generate diverse Fashion MNIST samples. Note that the unconditional score model using the loss in \cite{yansong} predicts $\bz/\sigma$. Proceeding as in Appendix A, we see that the corresponding energy (if it exists) may be viewed as a $\log(\cdot)$ function that saturates with $\bx$; the gradient of the function vanishes away from the minimum, which affects the convergence. The approach used to generate the samples in Fig. \ref{sampling_prior}.b avoids this by training a time-conditional model, which does not correspond to a single energy function. %On the other hand, the score network trained using \eqref{implicit} predicts $\bz \sigma$ whose integral would be a quadratic function with well-defined gradients even away from the minimum. Therefore, in this case, one does not need to tune the time scheduling hyper-parameter as conditioning of the score network is not required. 
These experiments show that the diversity and quality of samples generated by i-MuSE in Fig. \ref{sampling_prior}.a is comparable to that obtained from the traditional conditional score model. However, the i-MuSE formulation as a single energy function enables us to realize fast algorithms for MAP estimation with convergence guarantees.}
\mrpar{R2.C0\\R3.C2}{\subsection{Sampling from the posterior distribution}
In this section, we evaluate i-MuSE and diffusion model \cite{dps} under the context of sample generation from the posterior distribution.\\
1) Inpainting: We consider the inpainting problem, where we consider a box-type mask with added Gaussian noise with standard deviation $\eta= 0.01$. We consider three settings with increasing size of the masked region. Clearly, the problem becomes more difficult as larger regions of the image are masked. Fig. \ref{sampling_inpainting} compares the i-MuSE sampling performance with DPS for different inpainting masks. From Fig. \ref{sampling_inpainting}.a, we observe that when the mask covers only a small portion of the image (a), both DPS and i-MuSE are able to generate samples closer to the original image. However, when significant portions of the original image are masked, as in Fig. \ref{sampling_inpainting}.(b) \& (c), both the models generate samples that are not consistent with the original image. In this case, both models fill the masked region by sampling from the prior distribution. Since both models have comparable sampling performance, we consider only i-MuSE for the next sampling experiment.}
\\
\\
2) Reconstruction of MRI image: In this section, we consider sample generation from the posterior distribution in the context of MR image reconstruction. We considered sampling using Poisson 2D undersampling patterns, the generated samples showed low variability; these results are not shown in the paper.  We note that when the undersampling rate is low, the solution is unique with low variability. Therefore, we considered \mrpar{R4.C7}{6 and 8-fold 1-D Cartesian acceleration}, which are quite high.  Figs. \ref{sampling}.a and \ref{sampling}.b illustrate samples drawn using the Langevin MCMC algorithm, which is initialized randomly from a unit standard Gaussian distribution. We also show the estimated MMSE and the uncertainty map. The blurring in the samples is due to the high acceleration. From the figure, we observe that Langevin sampling is able to generate images with high variability (which are more visible in the enlarged images marked by arrows). This is also reflected in the variance map. In Fig. \ref{sampling}, we also report the negative log-prior (NLPr) and the negative log-posterior (NLPo), which are simply $\mathcal{I}_\theta(\bx)$ and $\dfrac{1}{2\zeta^{2}}\|\bA\bx-\bb\|_{2}^2 + \mathcal{I}_\theta(\bx)$, respectively. It should be noted that, unlike diffusion models, we do not have to evaluate complex integrals (see \eqref{LineInt_diffusion}) to compute these values. 
% We note that the NLPo of the samples generated by MM are similar irrespective of initialization, indicating that the solutions are valid MAP estimates.  % When the undersampling rate is low, the MAP estimate is unique. Therefore, the MM algorithm converges to the same solution (as seen in Fig. \ref{convergence}). However, at very high accelerations, the posterior may have multiple minima, and consequently, depending on the specific random initialization, the MM algorithm may converge to the minimum closest to the initialization. Since the MM approach converges to the MAP estimates, it does not offer an estimate of the spread or variance of each of the maxima. By contrast, Langevin sampling uses the stochastic differential equation, which adds Gaussian noise at each iteration. Thus, it offers samples that capture the spread of the distribution, even when the solution is unique.

\section{Discussion}
\mrpar{R4.C2}{The focus of this work is on learning the prior distribution from training data, followed by its use in MAP estimation. The proposed model may not readily apply to problems when there is a significant shift in the true prior distribution from the training data. For example, the energy model learned from the T1-weighted MR image data set may produce biased
reconstructions when used to recover a T2-weighted data set unless careful data augmentation strategies are used.}
\mrpar{R2.C6} {Furthermore, we note that the energy models are trained using DSM, which aims to eliminate Gaussian noise. Consequently, the model may sometimes be more effective in removing Gaussian noise than structured (alias) artifacts. See Fig. \ref{convergence}, where the e-MuSE scheme is observed to offer improved results with random initialization, compared to SENSE). }
\mrpar{R2.C5}{We note that the performance of E2E-MoDL can be further improved using DRUNet architecture as in i-MuSE. However, we note that using large networks within MoDL will translate to algorithms with high memory demand. }\\

\mrpar{R4. C4} {In the theory section, we modeled the noise $\bn$ as independent Gaussian white noise. The multichannel data acquired by the MR imaging scanner often have correlations between the coils, which we have not accounted for in the reconstructions. When the noise covariance matrix is available, this issue can be addressed by whitening the measured data. Since the noise covariance matrix is not available, we did not perform coil whitening. Consequently, the DC term in \eqref{eq:1} is an approximation. Correctly accounting for the noise properties can translate into improved performance and the ability to generalize the algorithms across different scanners and coils. We consider analyzing the impact of noise whitening as part of future work. }
\mrpar{R4.C8}{We note that the proposed scheme is
significantly more memory-efficient during
training, compared to unrolled algorithms. Thus, the proposed approach
may be beneficial in large-scale problems (e.g., 3D/4D),
where unrolled algorithms are infeasible. We anticipate that larger networks are required in such settings, where repeated evaluations of the corresponding score models during inference may translate to high computational complexity. We will explore these applications in our future work, when sufficient amount of data are available.}
\section{Conclusion}
We introduced multi-scale energy models, which can be used to derive the MAP estimate. The multi-scale strategy encourages the convergence of the algorithm, irrespective of the initialization, while ensuring estimation accuracy. We first introduced the e-MuSE model, which learns different energy networks corresponding to various smoothed versions of the true prior distribution. The e-MuSE energy is then used to define the negative log-posteriors at different scales, which are sequentially minimized using a MM-based algorithm to derive the MAP estimate. Although this approach offered improved convergence and accuracy over single-scale approaches, a challenge is the sensitivity to the specific choice of scales and exit conditions at each scale. We then introduced the i-MuSE model that learns a single energy, whose gradients at each scale are equivalent to the e-MuSE gradient at that scale. We showed that the i-MuSE energy of an image is a measure of the square of the distance from the data manifold. The smooth single i-MuSE energy function offers improved convergence and accuracy, while being independent of the schedule. Our results showed that i-MuSE offers MAP estimates that are comparable in quality to E2E trained MoDL, while being relatively insensitive to the forward model and initialization. Furthermore, the i-MuSE approach also allows us to sample from the posterior distribution, which can be used to determine uncertainty measures.

\section{Appendix}
\subsection{Proof of Proposition \ref{lemma2}}

We first show that when the dimension of the manifold is much smaller than the ambient dimension, the line joining $\bx$ and $\tilde \bx = \bx + \sigma \bz$ is orthogonal to the manifold $\mathcal M$ with high probability. 
\begin{lemma}
\label{lemma1}
Consider $\tilde{\bx} = \bx + \sigma \bz$ where $\bz \sim \mathcal N(0,\mathbf I)$. The energy of the projection of the vector $\boldsymbol{v}=\tilde {\bx}-\bx$ to the $d$-dimensional tangent space of the manifold $T_{\mathcal M}$ at $\bx$ satisfies:
\begin{eqnarray}
Pr\Big(\left\|P_{T_{\mathcal M}(\bx)}(\boldsymbol{v})\right\|^2 > \frac{(\alpha+\epsilon\sqrt\alpha)}{1-\epsilon}~\left\|\boldsymbol{v}\right\|^2 \Big) \leq \frac{4}{\epsilon^2 m},
\end{eqnarray}
for all  $\bx \in {T_{\mathcal M}}(\bx)$. Here $\alpha = d/m$, $\epsilon>0$.
\end{lemma}
\begin{IEEEproof}
  We consider $\|\tilde {\bx}-\bx\|^2 = \sigma^2 \|\bz\|^2; \bz \in \mathcal N(0,\mathbf I_m)$. \mrpar{R4.C10}{Using the properties of Gaussian random variables, it can be shown that the mean energy of the random variable $\boldsymbol s = \|\bz\|^2$ is $\mathbb E[\boldsymbol s]=m$ and its variance $\mathbb E[(\boldsymbol s-m)^2]=2m$. We have 
   \begin{eqnarray}\nonumber
       Pr(|\bs-m|>m\epsilon) &=& Pr(|\bs-m|^2>m^2\epsilon^2)\\\nonumber
       &\leq& \frac{\mathbb E[|\bs-m|^2]}{m^2 \epsilon^2} = \frac{2m}{m^2\epsilon^2} =\frac{2}{m\epsilon^2},\\
   \end{eqnarray}
   where we used Markov's inequality in the second step and $\epsilon >0$. We thus have:%randomness.pdf
   \begin{eqnarray}\nonumber
    Pr\Big(\|\bz\|^2<m(1-\epsilon_1)\Big) &<& Pr\Big(\Big|\|\bz\|^2-m\Big|>m\epsilon_1\Big)\\
    &\leq&   \frac{2}{m\epsilon_1^2}
   \end{eqnarray}
   where $\epsilon_1 >0$}. We now consider the coefficients of the projection of $\bz\sim  \mathcal N(0,\mathbf I_{d})$ to the d-dimensional orthonormal basis of $\mathcal T_{M}(\bx)$, denoted by $\boldsymbol{p}$. With similar arguments, we obtain 
   \begin{equation}
    Pr\Big( \|\boldsymbol p\|^2 > d(1+\epsilon_2)\Big) \leq \frac{2}{d\epsilon_2^2}
   \end{equation}
   where $\epsilon_1 >0$. Using the union bound, we obtain: \begin{equation}
    Pr\Big( \frac{\|\boldsymbol p\|^2}{\|\bz\|^2} > \frac{d(1+\epsilon_2)}{m(1-\epsilon_1)}\Big) \leq \frac{2}{d \epsilon_2^2}+\frac{2}{\epsilon_1^2 m}
   \end{equation}

   We set $\alpha = d/m$ and $\alpha \epsilon_2^2 = \epsilon_1^2 = \epsilon^2$. With this, we have: 
   \begin{equation}
    Pr\Big( \frac{\|\boldsymbol p\|^2}{\|\bz\|^2} > \frac{ (\alpha+\epsilon{\sqrt\alpha})}{(1-\epsilon)}\Big) \leq \frac{2}{m} \left( \frac{1}{\epsilon^2}+ \frac{1}{\epsilon^2}\right) = \frac{4}{m\epsilon^2}
   \end{equation}
\end{IEEEproof}
Note that as $\alpha = d/m \rightarrow 0$, the energy of the projection approaches zero with high probability. For example, when $m=10^6$, $d=2$ and $\epsilon = 0.01$, we have:  
\begin{equation}
    Pr\Big( \left\|P_{T_{\mathcal M}(x)}(\boldsymbol{v})\right\|^2 > 1.6\times 10^{-5} {\|\boldsymbol{v}\|^2}\Big) \leq  4\times 10^{-2}
   \end{equation}
Thus, the distance $d(\tilde{\bx},\mathcal M)=\|\tilde {\bx}-\bx\|=\|\boldsymbol{v}\|$ is the distance of $\tilde{\bx}$ to the manifold. 

With this, we now provide the proof for Proposition \ref{lemma2}.
\begin{IEEEproof}
When $\tilde {\bx}-\bx$ is orthogonal to the manifold, $\mathcal{I}_\theta(\bx)=0$ is the closest point to $\tilde{\bx}$ on the manifold. The distance of $\tilde {\bx}$ to the manifold is therefore given by: 
\begin{equation}
d(\tilde{\bx},\mathcal M) = \|\tilde{\bx} - \bx\| = \sigma \|{\boldsymbol z}\|.
\end{equation}
    We compute the i-MuSE energy at $\tilde{\bx}$:
       \begin{eqnarray}
        \mathcal{I}_\theta(\tilde \bx) &=& \underbrace{\mathcal{I}_\theta(\bx)}_0 + \int_{0}^{\sigma} \left\langle\underbrace{\nabla  \mathcal{I}_\theta(\bx + \sigma \boldsymbol z)}_{\sigma \boldsymbol z},\boldsymbol z\right\rangle d\sigma\\
        &=& \underbrace{\mathcal{I}_\theta(\bx)}_0 + \|\boldsymbol z\|^2\int_{0}^{\sigma} \sigma d\sigma =  \frac{1}{2} \|\sigma\boldsymbol z\|^2\\
        &\mrpar{R4.C11}{=}& \frac{1}{2} \|\tilde \bx-\bx\|^2\\
        &=&\frac{1}{2}~d(\tilde{\bx},\mathcal M)^2
    \end{eqnarray}
\end{IEEEproof}
The last line is due to Lemma \ref{lemma1}.
\subsection{MM algorithm for i-MuSE with data guidance}
We assume that the mapping $\hat{\bx}= \bx - \nabla_\bx ~I_{\theta}(\bx)$ is Lipschitz continuous with a constant $\beta$. This ensures that the Jacobian of the above mapping satisfies $\mathbf J_{\hat{\bx}} \leq \beta ~\mathbf I$. We now consider the Hessian of the data-term in \eqref{imusemod}:
\begin{equation}
\nabla_\bx^2 (\|\bA \hat{\bx}-\bb\|_{2}^2)=  \mathbf J_{\hat{\bx}}^T \left(\bA^T\bA\right)~ \mathbf J_{\hat{\bx}}
\end{equation}
The operator norm of the Hessian $
\nabla_\bx^2 (\|\bA \hat{\bx}-\bb\|_{2}^2)$ satisfies
\begin{eqnarray}\nonumber
\|\nabla_\bx^2 (\|\bA \hat{\bx}-\bb\|_{2}^2)\|&\leq&  \|\mathbf J_{\hat{\bx}}\|^2 \left\|\bA^T\bA\right\|~\leq~ \beta^2 \left\|\bA^T\bA\right\|,
\end{eqnarray}
which gives:
\begin{equation}\label{H_inq}
\nabla_\bx^2 (\|\bA \hat{\bx}-\bb\|_{2}^2)\preccurlyeq \beta^2~ \bA^T\bA
\end{equation}

% To prove \eqref{H_inq}, it is sufficient to show that 
% \begin{eqnarray}
%     \mathbf Q^T \bA^T \bA\mathbf Q &\leq& \alpha^2~ \mathbf A^T\mathbf A-\mathbf S
% \end{eqnarray}

% \begin{eqnarray*}
%     \|\mathbf Q^T \bA^T \bA\mathbf Q\| &\leq& \| \alpha^2~ \mathbf A^T\mathbf A-\mathbf S\|\\
%     &\leq &(\alpha^2  +1)\; \|\bA^T \bA\| + \|\mathbf Q^T \bA^T \bA\| + \|\bA^T \bA \mathbf Q \| \\
%     &\leq & (1+\alpha)^2~\|\bA^T \bA\|
% \end{eqnarray*}
% and this means that 
% $$\mathbf Q^T \bA^T \bA\mathbf Q \leq (1+\alpha)^2~\|\bA^T \bA\| I$$
% Because $\|\mathbf Q^T \bA^T \bA\mathbf Q\| \leq \alpha^2~ \bA^T \bA$  and since $\alpha >0$, the above condition is true. 

Using the above result and the second-order Taylor series expansion \cite{mm_tutorial}, the data-term can be upper-bounded as:
\begin{eqnarray}\nonumber
    \|\bA \hat{\bx}-\bb\|_{2}^2 
    &\leq& c_n + (\bx-\bx_n)^H \mathbf g_n \\&& \qquad \nonumber +\dfrac{\beta^2}{2} (\bx-\bx_n)^H  \bA^T \bA (\bx-\bx_n),\\
\end{eqnarray}
where $\mathbf g_n = \nabla_{\bx_n} (\|\bA \hat{\bx}_n-\bb\|_{2}^2)$ and $c_n = \|\bA \hat{\bx}_n-\bb\|_{2}^2$ is a constant.
We use Lemma \ref{lemma0}  to upper-bound the regularizer $\mathcal{I}_\theta(\bx)$. Therefore, the entire cost function in \eqref{imusemod} is upper bounded by the following surrogate function: 
\begin{eqnarray}\nonumber
    g(\bx|\bx_n) &=& \dfrac{1}{2}\bx^H\left(\dfrac{\beta \bA^T\bA}{2\zeta^2}+ {L\bI}\right)\bx -\\\nonumber
   &&\bx^H\bigg(\dfrac{\beta \bA^T\bA \bx_n-\mathbf g_n}{2\zeta^2}
    + L\bx_n-\nabla_{\bx_n} \mathcal{I}_\theta(\bx_n)\bigg)\\
\end{eqnarray}
The minimization of the above surrogate function has a closed-form solution and is given as in \eqref{sol_mod_dc}.

\bibliographystyle{IEEEtran}
\bibliography{ref}

\end{document}